\documentclass[iop]{emulateapj}
\usepackage{graphicx-psmin}
\begin{document}
%new commands and definitions
\newcommand{\kms}        {km~s$^{-1}$}
\newcommand{\herschel} {\textit{Herschel}}
\newcommand{\den}{${n}_{\mathrm{H}_2}$}
\newcommand{\ncolone}{$N_{\mathrm{CO}}$}
\newcommand{\tkin}{$T_{\mathrm{kin}}$}
\newcommand{\cmtwo}{cm$^{-2}$}
\newcommand{\cmthree}{cm$^{-3}$}
\newcommand{\coone}{$^{12}$CO}
\newcommand{\msun}{M$_{\odot}$}
\newcommand{\lsun}{L$_{\odot}$}
\newcommand{\mic}{$\mu$m}
\newcommand{\htwo}{H$_{2}$}
\newcommand{\lfir} {$L_{\mathrm{FIR}}$}
\newcommand{\lco} {$L_{\mathrm{CO}}$}
\newcommand{\jyb}{Jy beam$^{-1}$}
\def\msol{\ifmmode {\>M_\odot}\else {$M_\odot$}\fi}
\def\lsol{\ifmmode {\>L_\odot}\else {$L_\odot$}\fi}

\def\be{\begin{equation}}
\def\ee{\end{equation}}
\def\bdm{\begin{displaymath}}
\def\edm{\end{displaymath}}
\def\bea{\begin{eqnarray}}
\def\eea{\end{eqnarray}}

\title{Morphology and Kinematics of Warm Molecular Gas in the Nuclear
  Region of Arp~220 as Revealed by ALMA} 

\author{Naseem Rangwala\altaffilmark{1,2}, Philip R. Maloney
\altaffilmark{1}, Christine D. Wilson\altaffilmark{3}, Jason
Glenn\altaffilmark{1}, Julia Kamenetzky\altaffilmark{1}, and Luigi
Spinoglio\altaffilmark{4}} 

\altaffiltext{1}{Center for Astrophysics and Space Astronomy,
  University of Colorado, 1255 38th street, Boulder, CO 80303}  
\altaffiltext{2}{Visiting Scientist, Space Science and Astrobiology
  Division, NASA Ames Research Center, Moffet Field, CA 94035}  
\altaffiltext{3}{Dept. of Physics \& Astronomy, McMaster University,
  Hamilton, Ontario, L8S 4M1, Canada } 
\altaffiltext{4}{Istituto di Astrofisica e Planetologia Spaziali, INAF, Via Fosso del Cavaliere 100, I-00133 Roma, Italy} 
\begin{abstract}

We present Atacama Large Millimeter Array (ALMA) Cycle-0 observations
of the CO $J = 6-5$ line in the advanced galaxy merger Arp~220. This line
traces warm molecular gas, which dominates the total CO luminosity.
The CO emission from the two nuclei is well resolved by the
$0.39\arcsec \times 0.22\arcsec$ beam and the exceptional sensitivity
and spatial/spectral resolution reveal new complex features in the
morphology and kinematics of the warm gas. The line profiles are
asymmetric between the red and blue sides of the nuclear disks and the
peak of the line emission is offset from the peak of the continuum
emission in both nuclei by about 100 pc in the same direction. CO
self-absorption is detected at the centers of both nuclei but it is
much deeper in the eastern nucleus. We also clearly detect strong,
highly redshifted CO absorption located near the southwest side of
each nucleus. For the eastern nucleus, we reproduce the major line
profile features with a simple kinematic model of a highly turbulent,
rotating disk with a substantial line center optical depth and a large
gradient in the excitation temperature. The red/blue asymmetries and
line-to-continuum offset are likely produced by absorption of the blue
(SW) sides of the two nuclei by blue-shifted, foreground molecular
gas; the mass of the absorber is comparable to the nuclear warm gas
mass ($\sim$$10^{8}$ \msun). We measure an unusually high $L_{\mathrm{CO}}/L_\mathrm{FIR}$
ratio in the eastern nucleus, suggesting there is an additional energy
source, such as mechanical energy from shocks, present in this
nucleus.

\end{abstract}


\section{Introduction}
As the closest example ($D = 77 $ Mpc, 1\arcsec = 373 pc) of an
ultra-luminous infrared galaxy (ULIRG), Arp~220 has been studied
intensively at a wide range of wavelengths. It is an advanced merger
of two gas-rich galaxies with high dust optical depth ($\tau_{100\mu
m} \sim 5$; Rangwala et al. 2011) and a large reservoir of molecular
gas ($\sim 10^{10}$ \msun) concentrated in its nuclear region
\citep{scoville97}. It hosts one of the most extreme star-forming
environments in the local universe and is a widely used template for
interpreting observations of high-$z$, dusty star-forming
galaxies. High-resolution ground-based radio interferometric
observations show that it has two nuclei separated by roughly
1\arcsec\ \citep[e.g.,][]{scoville97, sakamoto09}, with evidence of a
highly dust-obscured AGN in its western nucleus \citep{downes07,
engel11}. Additionally, a kiloparsec-scale central molecular disk
surrounding the two nuclear disks has been inferred from CO $J = 2-1$
interferometric observations by \citet{scoville97}.

Observations of high-$J$ CO rotational transitions from the
\textit{Herschel Space Observatory} of nearby galaxies (including
Arp~220) and Galactic star-forming regions have determined that there are generically two components in
the molecular gas: a warm (high-pressure) component traced by
mid-to-high-$J$ lines that comprises a minority of the mass ($\sim$10\%
of the total molecular gas mass) but dominates the CO luminosity
($\sim$90\% of the total luminosity), and a cold (low-pressure)
component traced by the lowest-$J$ lines, which dominates the mass
\citep[e.g.,][]{panuzzo2010, rangwala2011, kamenetzky12, spinoglio12,
meijerink13, goicoechea13, etxaluze13, pereira-santaella13,
papadopoulos14}. Comparison of the observed CO spectral line energy
distributions (SLEDs) and far-infrared luminosities to theoretical
models indicates that in several prominent cases (e.g., Arp 220, M 82,
NGC 6240, M83) the gas heating is dominated by some mechanism other
than UV or X-ray photons (as in photon-dominated and X-ray dominated
regions) or cosmic rays; mechanical energy from shocks generated by
supernovae and stellar winds are the most plausible sources. A recent
study by Greve et al. (2014) using CO measurements in 62 local
(U)LIRGs and 35 high-$z$ galaxies also concludes that mechanical
energy is the more likely source of energy for this warm component.
The pressure in the warm component is found to be typically two orders
of magnitude larger than the cool component pressure (e.g., $10^7$ K
cm$^{-3}$ vs. $10^{5}$ K cm$^{-3}$).  These findings suggest that this
warm component of the molecular gas is intimately associated with the
feedback processes, which play an essential role in galaxy formation
and evolution.

Our ongoing \textit{Herschel} archival survey is systematically
deriving physical conditions of the molecular gas in a large sample
($\sim$200) of nearby galaxies by modeling the combined observations
from the \herschel\ SPIRE-FTS instrument, which provides CO SLEDs from $J = 4-3$ to $J = 13-12$, and
ground-based measurements of low-$J$ CO lines. The results from the
analysis of an initial sample of 17 galaxies \citep{kamenetzky14}
confirmed the main findings described above. 
This study also found that the mid-$J$ CO transitions ($J > 4-3$), where
the CO spectral line energy distribution peaks, primarily arise in the
warm component and their line luminosities are well correlated with
the total CO luminosity. Thus these transitions can be used as a
reliable tracer of warm molecular emission. Because of its relatively large beams,
\textit{Herschel} could not spatially resolve most galaxies, so only
average properties of the warm gas have been inferred, and it has been
difficult to associate the warm gas with the distribution of star
formation, cool molecular gas, and nuclear activity.

The CO $J=6-5$ line is now accessible with ALMA, which can finely
resolve the morphology and kinematics of the warm gas in many nearby
galaxies. This transition has been previously observed in Arp 220 with
the SMA \citep{matsushita09}. However, the lesser resolution and lower
sensitivity of SMA at these higher frequencies could not resolve the
emission from the two nuclei. In this paper, we present the first
high-resolution and high-S/N maps of the warm molecular gas in Arp 220
using the CO $J=6-5$ line. The corresponding continuum observations (at
435\mic) were presented in \citet[][Paper~I hereafter]{wilson14}.
These observations showed that the western nucleus is considerably
warmer and more optically thick
($T_\mathrm{dust} = 200$ K$, \tau_{434\mu \mathrm{m}}$ = 5.3) than the eastern nucleus ($T_\mathrm{dust} = 80~$K$, \tau_{434\mu\mathrm{m}}$ = 1.7). The two nuclei combined account
for roughly 80\% of the total infrared luminosity of Arp~220.

The ALMA observations and data reduction are presented in Section 2. The observed morphology and kinematics in the CO~$J=6-5$ emission is described in Section 3 followed by detailed modeling of the kinematics of the nuclear disks in Section 4. We discuss disk stability and the energetics of the warm molecular gas in Sections 5 and 6, respectively. Detection of absorption in the SiO $J=16-15$ is presented in Section 7 and our conclusions are summarized in Section 8.

\section{Observations and Data Reduction}
Arp 220 was observed with ALMA on 2012 December 31 (Project
2011.0.00403.S) at the end of cycle-0 in an extended/hybrid
configuration. The Band 9 receivers were tuned to cover the CO J=6-5
transition (rest frequency 691.473 GHz; Arp 220's redshift = 0.01813)
and adjacent continuum, with a total bandwidth of 8 GHz; however, the
usable bandwidth was somewhat reduced due to overlap between the
spectral windows covering the CO line and the need to flag a handful
of end channels in each of the four spectral windows. The observations
covered projected baselines from 13 to 374 m. The correlator was
configured in low-resolution wide-bandwidth TDM mode to give a
spectral resolution of 15.625 MHz. The total on-source integration
time was 44.4 minutes using 23 antennas. 

The data were flagged,
calibrated, %(including phase-only self calibration), 
and imaged by ALMA
staff prior to delivery. 
%For this analysis, we used the calibrated
%$uv$ data as delivered to us but improve the imaging and
%self-calibration as described below.
The data have two unusual technical problems. First, no absolute flux
calibrator was observed as part of the data set. The absolute flux
scale was set using a flux of 6.0 Jy for the bandpass calibrator,
3C279, determined by a Band 9 observation taken within one day of
these observations. This process introduces some additional
uncertainty on the absolute flux scale which is difficult to
quantify. 
Second, the signal-to-noise ratio on the phase calibrators
is very low, which required a major adjustment 
to the standard cycle-0
calibration process. The process used by ALMA staff was as
follows. (1) Obtain a bandpass calibration solution from the
bandpass calibrator. (2) Use the bandpass calibrator to
solve for phase offsets among the four spectral windows and two
polarizations. (3) Obtain a solution for phase versus time
from the bandpass and two gain calibrators while combining the data for both
polarizations and all four spectral windows together. (4) Obtain a
solution for amplitude versus time from the three calibrators again
combining the data from all polarizations and spectral windows. Use
this solution to estimate the fluxes (0.32, 0.40 Jy) 
of the two secondary calibrators. (5)
Obtain a final solution for amplitude versus time using the adopted
flux of the two gain calibrators.

For our analysis, we began with the delivered calibrated
$uv$ data and made new, cleaned data cubes following the same steps
carried out by ALMA staff using CASA version 4.1.0. Repeating the imaging allowed us to adjust
the clean boxes to reduce the level of artifacts in the final cubes
even further from the very good images originally delivered to us and
to improve the resolution by switching to uniform weighting. We
also imaged all four spectral windows and adjusted the velocity ranges
used to make the continuum images \citep{wilson14} to avoid all
emission and absorption features.

The reduction notes from the ALMA staff noted
that self-calibration was required to get the best images. 
Following their method, we carried out two rounds of 
phase-only self-calibration using natural weighting and a
relatively small clean box to include emission from the bright central
region. A final round of self-calibration on
both amplitude and phase significantly improved the residuals in the
data cube. As noted by ALMA staff, we see significant amplitude
corrections are required and these corrections correlate among
the different spectral windows.
These steps significantly improved the noise level in the cube, from
36 mJy/beam to 16 
mJy/beam in the line free channels and from 100 mJy/beam to 18
mJy/beam in channels where the line emission was strong. 

Before making the final image cube, we carried out continuum
subtraction using a zeroth order fit to most of the channels in the
two spectral windows located away from the bright CO emission.
In the final imaging step, the clean box was adjusted with 
frequency to include all regions with significant emission, which
improved the negative residuals in certain channels with bright
emission. We made image cubes with both natural and uniform weighting 
to maximize sensitivity and 
angular resolution, respectively. We used a velocity resolution of 10 km
s$^{-1}$ for the CO data cube. As a final step, the uniform cube was smoothed
slightly so that all channels would have the same beam ($0.39\arcsec
\times 0.22\arcsec$) at position angle 28$^{\circ}$.  The typical
1$\sigma$ noise in the cleaned uniform cube is 30 mJy beam$^{-1}$ or 0.9~K
per channel.

\begin{figure*}[ht]
\includegraphics[scale=0.52]{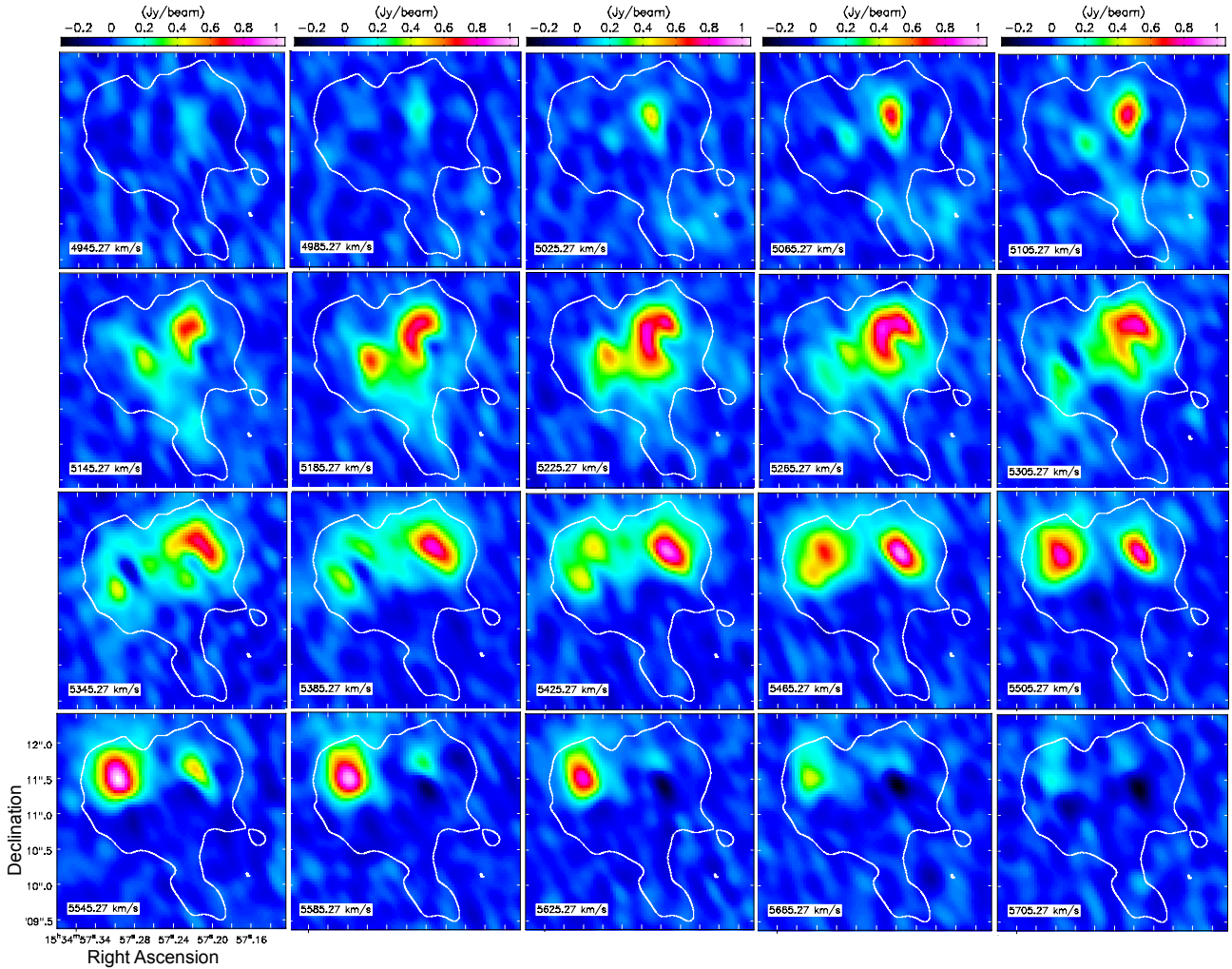}
 \caption{Channel maps covering the CO emission and absorption over $\sim$800 \kms. The 1\% contour level for the continuum emission is shown relative to the peak continuum flux of 1.38 \jyb. Redshifted CO absorption can be seen in the high velocity channels (5505 -- 5705 \kms). Note: The maps and spectra in this paper are not corrected for the small primary beam attenuation.} 
\label{chmaps}
\end{figure*}

\begin{figure*}[ht]
\center
\includegraphics[scale=0.45]{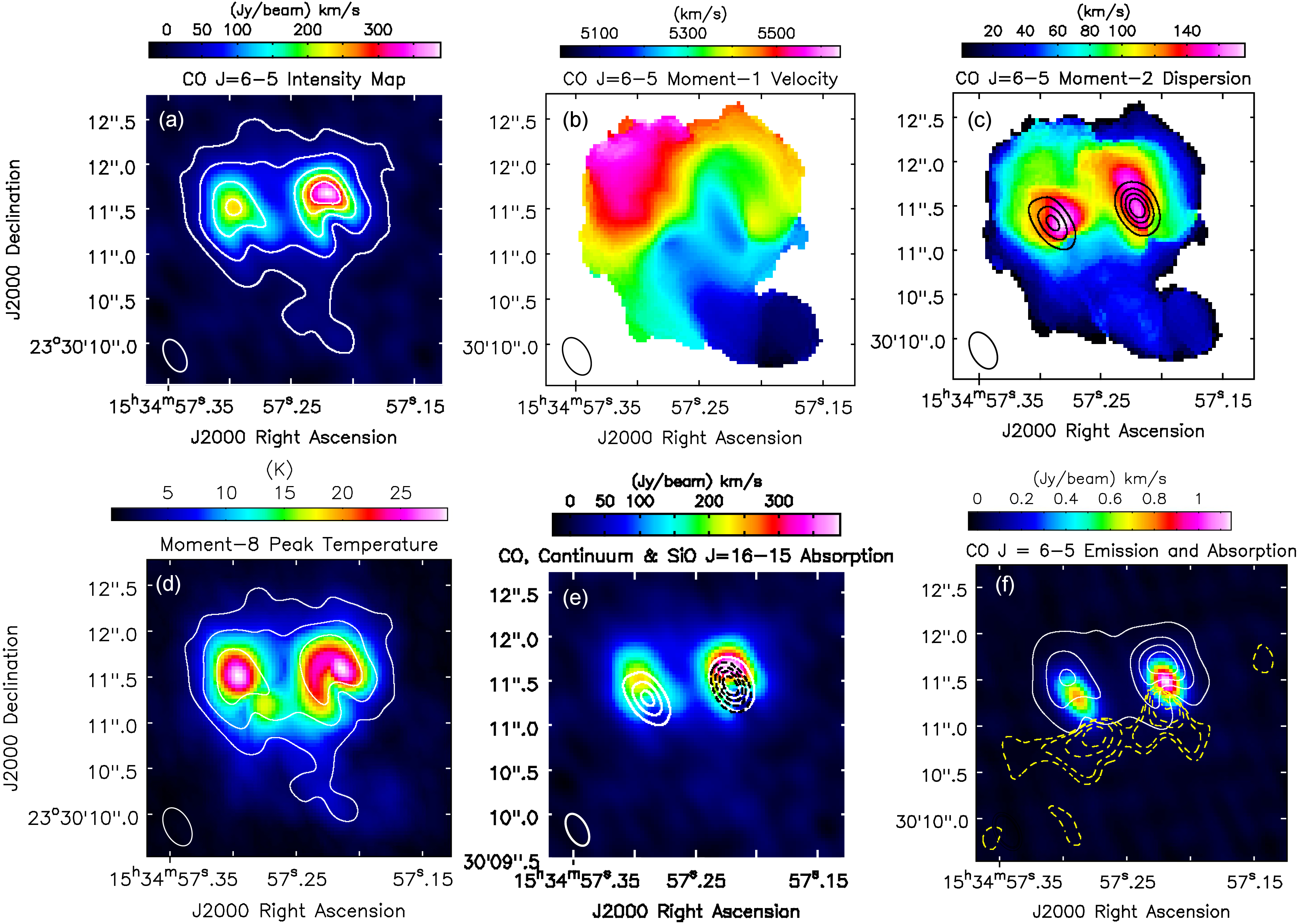}

\caption{ALMA high-resolution images of the CO $J = 6-5$ line and 691
GHz continuum in Arp 220.  (a) CO integrated intensity (moment-0)
map made from uniform cube. The 10\%, 20\%, 40\%, 60\% and 80\% contour levels are shown
relative to the peak flux of 389 Jy beam$^{-1}$ \kms\ in the western
nucleus. (b) Mean velocity Field (moment-1). (c) Velocity dispersion
(moment-2) map. The 20\%, 40\%, 60\% and 80\% continuum contours are
shown relative to the peak continuum flux of 1.15 \jyb\ in the
western nucleus. (d) Peak temperature (moment-8) map and integrated CO
intensity contours in White. (e) The peak of the line emission (color
map) is offset from the peak of the continuum emission (solid
contours) by roughly 100pc. The SiO $J=16-15$ line absorption (dashed
contours -- 20\%, 40\%, 60\% and 80\% of -70 \jyb\ \kms) is detected only
in the western nucleus and is unresolved. (f) Moment-0 contours of CO $J=6-5$ redshifted absorption over 5505 -- 5725 \kms\ (yellow contours). The continuum emission
(color) and line emission (solid contours) are overplotted. The yellow contour levels are 20\%, 30\%, 50\%, 70\% and 90\% relative to -25 \jyb\ \kms. The contours depict both the compact and extended components of the redshifted absorption.}
\label{comaps}
\end{figure*}

After a small correction for primary beam attenuation, the total integrated CO $J=6-5$ flux is 3670 Jy \kms. This value is in agreement with the total flux obtained from \textit{Herschel}-SPIRE of
$4070 \pm 80$ Jy \kms\ \citep{rangwala2011} within their calibration
uncertainties (about 10\% for Herschel-SPIRE and 15\% for ALMA).
Even in the absence of an absolute flux calibrator, the
final flux calibration of our data cube is consistent with literature values. 

The channel maps from the uniform data cube are presented in Figure \ref{chmaps} for CO emission over $\sim$800 \kms. The extent of the continuum emission is indicated by the 1\% contour level. Deep CO self-absorption is detected around the systemic velocity of the eastern nucleus that appears to go below zero. Also, redshifted CO absorption is seen in higher velocity channels.  These features are seen in the spectra at the continuum positions even before the continuum is subtracted. There are gaps in the data cube over two velocity intervals: 4245 -- 4305 \kms\ and 5735 -- 6035 \kms\ (see Figure \ref{velprofcen}). The latter gap is close to the redshifted CO absorption and limits our ability to measure the full velocity extent of this absorption feature. These features are further discussed in \S 3.

\begin{deluxetable*}{l|l|l|l}
\tablecaption{ALMA Arp 220 line Measurements}
\tablehead{\colhead{Quantity}& \colhead{Eastern Nucleus}&  \colhead{Western
    Nucleus} & \colhead{Units} } 

\startdata
\cutinhead{CO $J=6-5$ Emission}
Peak flux location  & (15:34:57.30, +23:30:11.53) & (15:34:57.22, +23:30:11.68) & J2000\\
 Integrated flux ($1\arcsec \times 1\arcsec$)   & $1070$ & $1305$  & Jy km s$^{-1}$ \\
 Peak flux  &  330  & 515  & Jy beam$^{-1}$ km s$^{-1}$\\
 Gaussian FWHM  &  0.83 $\times$ 0.63 & 0.66 $\times$ 0.63 & \arcsec \\
 Gaussian-fit Integrated flux  &  1565 &  1795  & Jy km s$^{-1}$ \\
 Line-to-Continuum ratio &  1035 & 990 & \kms \\
 $L_{\mathrm{CO}}/L_\mathrm{FIR}$\tablenotemark{a}  & $6.7^{+16}_{-4.4} \times 10^{-5}$ & $6.0^{+29}_{-3.6} \times 10^{-6}$  & \\
Deconvolved FWHM &  $0.74 \times 0.59$ & $0.59 \times 0.54$ & \arcsec \\
 Warm Gas Mass\tablenotemark{b} &  $1.4 \times 10^{8}$ & $1.6 \times 10^{8}$ & $M_{\odot}$\\
Warm Gas Column Density\tablenotemark{c} &  $9.2 \times 10^{22}$ & $1.5 \times 10^{23}$ & H/cm$^{-2}$ \\
 \cutinhead{SiO J = 16-15 Absorption}
Peak flux location &  No Detection & (15:34:57.22, +23:30:11.43) & J2000 \\
Peak flux &  -- & -80 & Jy beam$^{-1}$ km s$^{-1}$ \\
Gaussian FWHM, PA  &  -- & $0.385 \times 0.255$, 27 & ($\arcsec$,$^\circ$) \\
Gaussian-Fit Integrated flux &  -- & -87 & Jy km s$^{-1}$\\
 Mean depth &  -- & -0.23 & Jy beam$^{-1}$\\
Velocity extent & -- & $\gtrsim 400$ & km s$^{-1}$\\
\enddata

\tablecomments{The peak and integrated fluxes corrected for primary beam attenuation are reported from the
 natural weighted (higher sensitivity) moment-0 map with a beam size
 of $0.447\arcsec \times 0.273\arcsec$ ($\mathrm{PA} = 28^{\circ}$),
 while the source FWHM is reported from the higher-resolution uniform
 weighted map with a beam size of $0.39\arcsec \times 0.22\arcsec$.\\ 
The S/N in our data is very high -- $\sim$55 in the peak flux and $>
 400$ on the integrated flux -- leading to very small measurement
 errors; thus they are not reported. The overall uncertainties are
 dominated by systematic calibration uncertainties of $\sim$15\%.}  
\tablenotetext{a}{Using the \lfir\ measured from the ALMA 691 GHz
 continuum emission in Paper I. Note that the large range in
 uncertainty in this ratio arises entirely from the large systematic
 uncertainty on the \lfir. } 
\tablenotetext{b}{Using $M(H_2)_\mathrm{warm}/L_{CO6-5} = 21$
 \msun/\lsun\ derived from Herschel FTS CO SLED (Rangwala et
 al.\,2011)} 
\tablenotetext{c}{Using the Gaussian FWHM and $\mathrm{Area} = \pi\,R^2$}
\end{deluxetable*}

\section{Observed Morphology and Kinematics of the Warm Molecular Gas}

The CO $J=6-5$ emission traces the warm molecular gas peaks at the
centers of the eastern and western nuclei, which are unambiguously
resolved in the $0.39\arcsec \times 0.22 \arcsec$ resolution CO map
(Figure \ref{comaps}a), and has an extended component surrounding
the two nuclei. This overall morphology agrees with the CO
$J=2-1$ and CO $J=3-2$ maps of \citet{sakamoto99, sakamoto08}, in
which the emission associated with the two nuclei was characterized as
compact nuclear disks that are embedded in an extended outer disk. A simple
Gaussian fitting to the two nuclei shows that the western nucleus is
almost circular with a deconvolved size (FWHM) of $\sim 0.55\arcsec$
(205 pc using 1\arcsec = 373 pc), while the eastern nucleus is
elongated in the NE-SW direction with a deconvolved size (FWHM) of
$0.74\arcsec \times 0.59\arcsec$ (276 pc $\times$ 220 pc). This
indicates that the CO emission is extended compared to the dust
continuum emission, for which the half-maximum diameters are $76
\times (\leq 70)$ pc and 123 $\times$ 79 pc for the western and
eastern nucleus, respectively (Paper~I). However, there is likely
extended, low surface brightness dust emission that is below the
current sensitivity of our observations: the combined continuum
emission at 692 GHz detected in the two nuclei is about 80\% of the
total continuum emission measured in the ALMA map (Paper I) and only
56\% of the total continuum emission measured in the \textit{Herschel}
data \citep{rangwala2011}. The peak flux, integrated flux and source
sizes for CO $J=6-5$ emission are listed in Table 1.
 
The overall CO $J=6-5$ velocity field (Figure \ref{comaps}b) is also
similar to the CO $J=2-1$ and CO $J=3-2$ maps, which show a large
velocity gradient in the NE - SW direction. The velocity extent is
$\sim$500 \kms\ within 0.3\arcsec\ of the nuclei, 
 again similar to the
CO $J=2-1$ velocity extent \citep{sakamoto99}. The CO $J=6-5$
velocity dispersion ($\sigma = \mathrm{FWHM}/2.3548$) map (Figure
\ref{comaps}(c)) shows the dispersion peaks are coincident with
the continuum peaks, with values as large as 175 \kms\ and 160 \kms\
for the eastern and western nucleus, respectively.

To compare with the CO $J=2-1$ and CO $J=3-2$ maps from
\citet{sakamoto08}, which have a beam FWHM of 0.5\arcsec, we convolved
the CO $J=6-5$ line map to match their beam size. Using these maps we
find that about 50\% of the total detected CO $J=6-5$ emission arises
from within the nuclear region (i.e., within a radius of
0.45\arcsec). This fraction is higher than for the low-$J$ lines
($\sim$30\%), implying that the warm molecular emission is relatively
more compact.

\begin{figure}[ht]
\includegraphics[scale=0.27]{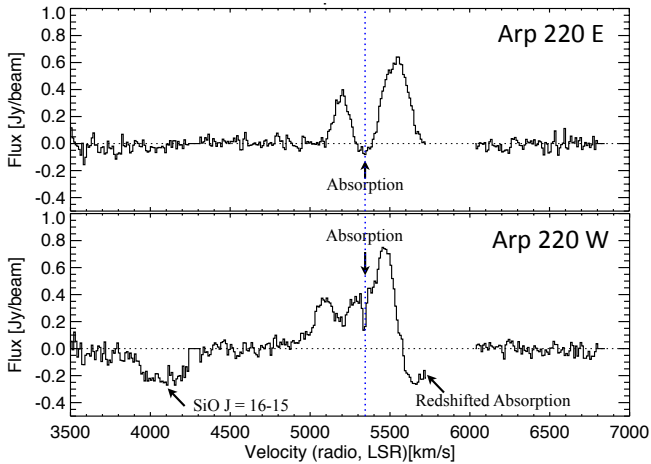}
\caption{Continuum subtracted spectral line profiles at the location
  of the continuum peak for the eastern and western nuclei, in the reference frame of CO.}
\label{velprofcen}
\end{figure}

The peak temperature (moment-8) map, Figure \ref{comaps}(d), displays
two noticeable features in comparison with the integrated flux map:
the semicircular ridge of emission in the western nucleus is more
prominent in peak temperature than in integrated flux, and there is an
off-nucleus ``hot spot" in the eastern nucleus (although its temperature is lower
than the nuclear peak). Comparable high-resolution maps in additional CO transitions will be required to
determine whether these temperature features correspond to variations
in excitation temperature.

Overplotting the continuum contours on the CO $J=6-5$ integrated
intensity map (Figure \ref{comaps}e) shows an offset between the
peak of the continuum emission and the peak of the line emission in
both nuclei; this offset is about 0.2\arcsec\ and has the same
direction. This offset is not seen in the low-$J$ lines (up to $J =
3-2$) from previous observations. We explain in Section 4 that this
offset is only apparent, and is likely caused by absorption of the entire SW
side of the line emission in the two nuclei.

Moreover, CO absorption is detected at the centers of the two nuclei
as shown in Figure \ref{velprofcen}; the absorption is much deeper in
the eastern nucleus and goes below zero in the continuum subtracted
spectra. This absorption is also shown in the channel map close to the
systemic velocity of the eastern nucleus (Figure \ref{chmaps}; channel 5345 \kms). 
The central dip seen in the eastern nucleus suggests the presence of absorption;
that this dip goes below zero (these are continuum subtracted spectra)
indicates both that this feature is the result of absorption and that
we are seeing absorption of the $6-5$ line against the continuum
source, i.e., the absorbing gas is absorbing both line and continuum
emission. 

\begin{figure}[ht]
\includegraphics[scale=0.4]{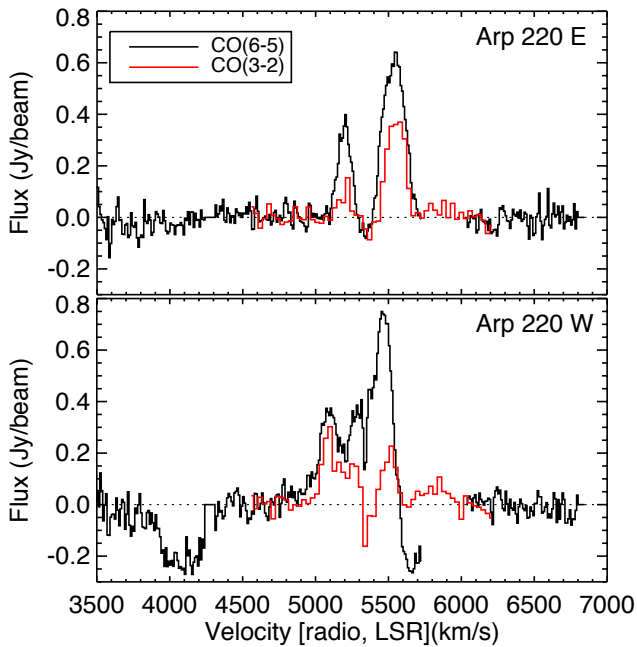}
%\vspace{-0.2in}
\caption{Comparison between the CO $J=6-5$ (this work) and CO $J=3-2$
  \citep[private communication 2015]{sakamoto09} velocity profiles at the location of the
  continuum peak for the eastern and western nuclei in $0.39\arcsec
  \times 0.22\arcsec$ and  0.3\arcsec\ beams, respectively.}  
\label{velproflowj}
\end{figure}

The deep CO absorption at the center was not reported from the interferometric
observations of the CO $J=2-1$ and $3-2$ lines with 0.5\arcsec\
resolution \citep{sakamoto08}. In Figure \ref{velproflowj}, the CO $J=3-2$ profile from
\citet[private communication 2015]{sakamoto09}, observed using a 0.3\arcsec\ beam, is overplotted
on our CO $J=6-5$ profile (at the continuum peak). This comparison
clearly shows that for the eastern nucleus a similar absorption dip is
present in the low-$J$ CO line, although it was not discussed by
\citet{sakamoto09} (further discussion is provided in section 4). In
fact, the CO $J=3-2$ line shows a double-peaked profile in the eastern
nucleus similar to that of CO $J=6-5$, except the blue side is much
weaker than the red side. Comparison of the western nucleus profiles also shows
similarity between the two transitions, i.e., a central absorption dip
and an overall complexity in the line profiles.

The spectrum of the western nucleus in Figure \ref{velprofcen} (bottom
panel) also shows redshifted CO absorption at velocities between
roughly 5505 \kms\ and 5725 \kms, much higher than the systemic
velocity of the system. Absorption is also detected towards the eastern nucleus but in off-center positions and in higher velocity channels (see Figure \ref{chmaps} (channel 5705 \kms) and Figure \ref{velprofs}). The redshifted absorption is shown with yellow
contours in Figure \ref{comaps}(f), which also shows the CO emission
in solid contours and the continuum emission in color. The absorption has a deep, compact component as well as a shallow, extended component. We note that although the shallow extended component is present over many velocity channels, the less than ideal calibration of the data and the low S/N mean that we cannot be certain it is not an artifact of the cleaning process.
The peak of the redshifted CO absorption is also offset in the two nuclei relative to
the continuum peak but in the opposite direction compared to the line
emission peak, i.e., towards the SW direction. 
Because this absorption is present towards both nuclei, its plausible that it arises from a continuous feature covering the southern part of the whole system, i.e., at least 400 pc in
length (roughly the distance between the two nuclei). This continuous
redshifted absorption feature could be interpreted as an infalling
molecular filament (see section 4.2), not unexpected in a chaotic
late-stage merger. It is difficult to ascertain the full velocity
extent of this feature because of gaps in our data but it is at least
200 \kms\ wide. 

A highly excited SiO line, $J = 16-15$, is detected in absorption
(dashed contour in Figure \ref{comaps}(e); also see Figure
\ref{velprofcen}), but only towards the western nucleus. This line is
discussed further in Section 7.

In summary, the exceptional sensitivity and spectral resolution from
ALMA has allowed us to detect several components in the warm molecular gas
in Arp 220 showing more complexity in the morphology and nuclear
kinematics than previously observed. 

\section{Modeling: Kinematics of the Nuclear Disks} 
\begin{figure*}[ht]
\center
\includegraphics[scale=0.43]{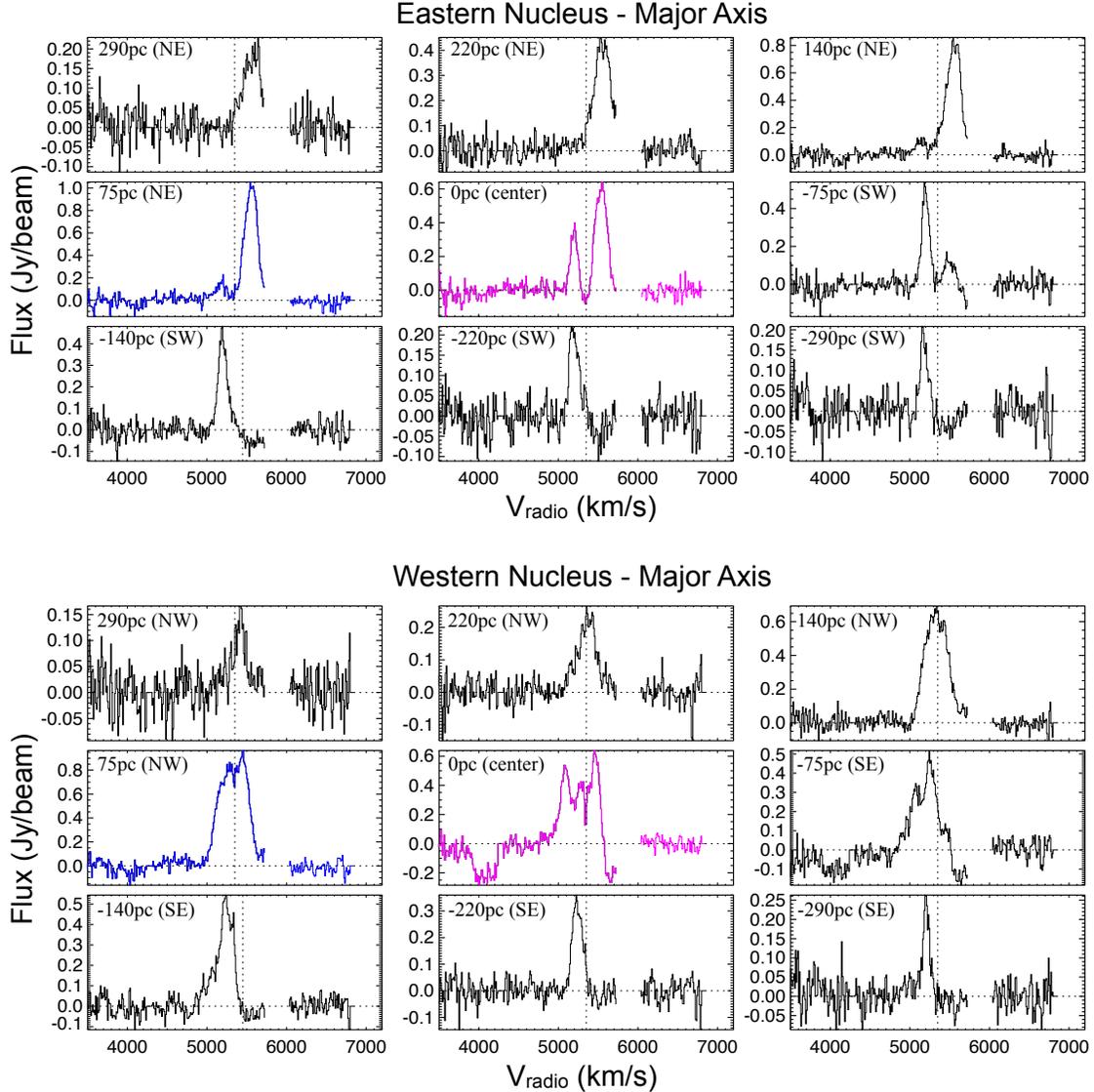}
%\vspace{-0.2in}
\caption{The velocity profiles along the major axes of the Eastern (for PA $\sim$35$^{\circ}$) and
  Western (for PA $\sim$111$^{\circ}$) nuclei. The blue and magenta colored profiles are at the
  location of the line peaks and continuum peaks, respectively. The vertical line marks the systemic velocity of the system at 5350 \kms.
}     
\label{velprofs}
\end{figure*}

Both the eastern and western nuclei show clear kinematic evidence for
rotating disks. In addition to the velocity gradients clearly seen in
Figure \ref{comaps}(b), in Figure \ref{velprofs}, we show spectra
along the major axes of both nuclear disks, spaced at multiples of a
half-beamwidth ($\sim$80 pc) from the dust continuum peaks. In both
cases we see clear signs of rotation: a systematic shift in the velocity of the emission
peak with position along the major axis, with an isolated red
or blue peak on opposite sides of the continuum peak. The kinematics
of the eastern nucleus, in particular, appear relatively simple, with
a double-peaked (but asymmetric) profile at the continuum center that
shifts to single red- or blue-shifted peaks with increasing distance
along the major axis. In comparison, the velocity profiles of the
western nucleus are more complex. For this reason we present here a
simple toy model intended to explain the main features of the spectra
for the eastern nuclear disk, and then discuss the implications for
both nuclei and the larger-scale gas distribution around the nuclei.

In the presence of velocity dispersion and systematic motion (e.g.,
rotation or outflow), we can write the line optical depth as a
function of frequency as
\be
\tau_\nu = \tau_0\phi_\nu
\ee
where $\tau_0$ is the line center optical depth and the Doppler line
profile (for non-relativistic velocities) is given by
\be
\phi_\nu = \exp\left({-\left[ {\nu - \nu_0\over\Delta\nu_D} -
    {(\nu_0/c){\bf n\cdot v}\over\Delta\nu_D}\right]^2}\right)
\label{phi_line}
\ee
where $\Delta\nu_D$ is the Doppler width ($\sqrt 2$ times the
dispersion), $\nu_0$ is the line center frequency, ${\bf n}$ is the
direction of propagation, and ${\bf v}$ is the velocity
\citep[e.g.,][]{hr68}. 

The nuclear disk is substantially inclined to our line of sight, with
$i\approx 50^\circ$ based on the deconvolved continuum source size
(Paper I). For simplicity, we assume that the disk is perfectly
edge-on. We also assume that the disk is uniform in the vertical
direction, which means that the disk thickness is only a meaningless
scaling parameter. This reduces the model to two dimensions, which
considerably speeds up the numerical modeling. We expect that the
inferred model parameters would only change by factors of order unity
in going from this edge-on, 2D disk to a tilted, 3D disk, e.g., the
true disk rotation curve will increase by $1/\cos i$ from the model
curve. The only features not captured by this approximation would be
systematic motions that are confined largely or entirely to the disk
vertical direction. While our simple edge-on model can only include features that are a function of projected distance from the disk center with no dependence on angle within or height above the disk, the match between the model spectra and the data are sufficiently good (see §4.2) that features violating this restriction are not required to explain the observed spectra.

We choose a coordinate system such that the $x$-axis lies in the plane
of the sky, the $y$-axis is now the disk vertical direction, and the
$z$-axis lies in the disk and is oriented along the line of sight
towards the observer, with $\theta$ the angle in the disk with respect
to the $x$-axis (so $\theta=\pi/2$ points towards the observer). The
radial coordinate $r$ then lies in the plane of the disk, and the
distinction between cylindrical and spherical radius has been removed
(e.g., a spherical outflow lies purely in the disk plane in this
approximation). The projected line-of-sight velocity is then 
\be 
{\bf n\cdot v} = V_c(r)\cos\theta + V_{out}(r)\sin\theta
\label{vproj}
\ee 
for circular rotation velocity $V_c$ and purely radial outflow
velocity $V_{out}$. The distortion of the line profile function by the
systematic velocity terms plays a key role in determining the
appearance of the disk emission. Note that equation (\ref{phi_line})
with equation (\ref{vproj}) is antisymmetric about $\theta=\pi/2$ in the
rotation-only case, and symmetric in the outflow-only case (provided
that the outflow is itself symmetric about this line), which cuts the
needed amount of computation in half. However, it is neither in the case
that both rotation and outflow are present.

We specify the line and continuum source functions and optical depths
as a function of radius in the disk, which is assumed to be circular
and axisymmetric. Both the source functions and the optical depth
profiles are assumed to be Gaussians in radius $r$.\footnote{Note that
this is not a {\it physical} model, in the sense that the excitation
of the CO line has not been calculated from conditions in the disk:
the line source function and optical depth are specified independently
of one another, and both the line and continuum source functions have
arbitrary normalizations. This model is intended to aid in
understanding the main features of the emission from the eastern
nucleus, not reproduce it in detail.} With both line and continuum
emission and absorption, the source function $S_{\nu}$ is given by the ratio of
the combined emissivities $j_{\nu}$ to the sum of the absorption coefficients $\alpha_{\nu}$:
\be S_\nu = {j_{\nu,l} + j_{\nu,d}\over {\alpha_{\nu,l} + \alpha_{\nu,
d}}}
\label{source1}
\ee
with the $l$ and $d$ subscripts referring to line and dust,
respectively. The optical depth at frequency $\nu$ is now 
\be
\tau_\nu = \tau_{\nu,l} + \tau_{\nu,d}.
\ee
In terms of the line-only and dust-only source functions, equation
(\ref{source1}) can be written 
\be 
S_\nu = {\alpha_{\nu,l}S_{\nu,l} +
  \alpha_{\nu,d}S_{\nu,d} \over {\alpha_{\nu,l} +
  \alpha_{\nu,d}}}
\label{source2}
\ee
i.e., the individual source functions weighted by their respective
absorption coefficients, or, equivalently, by the corresponding values
of $\tau_\nu$. Equations (\ref{source1}) and (\ref{source2}) can be
simplified by noting that we can take the dust emissivity and
absorption coefficient (and thus the dust source function) to be
independent of frequency over the width of the line profile.

The emitted intensity along a line of sight through the disk is then
\be
I_\nu = \int_{z_{min}}^{z_{max}} S_\nu\exp\left[-\int_z^{z_{max}}
  \alpha_\nu(z')dz'\right] \alpha_\nu dz\;.
\ee
To solve this equation, we divide the disk into segments along the
$z-$direction at each value of impact parameter $p$ (perpendicular
distance from disk center); changing variables from distance along the
line of sight to optical depth $\tau_\nu(z) = \int_z^{z_{max}}
\alpha_\nu(z')dz'$, $I_\nu$ is then the sum over pieces of the form
\be
S_\nu(z)(1-e^{-\tau_\nu(z)})e^{-\tau_\nu^{tot}(z)}
\ee
where $\tau_\nu(z)$ is the local optical depth, and
$\tau_\nu^{tot}(z)$ is the total optical depth from depth $z$ to the
surface of the disk. Hence
\be
I_\nu = \sum_{z_{min}}^{z_{max}}
S_\nu(z)\left(1-e^{-\tau_\nu(z)}\right) e^{-\tau_\nu^{tot}(z)}
\label{inu_eqn}
\ee
The model line profiles are calculated by convolving the disk emission
with a gaussian beam, with the model beam size matched to the
observations.

\subsection{Model Results - Axisymmetric rotating disk} 
\begin{figure}[ht]
\includegraphics[scale=0.37]{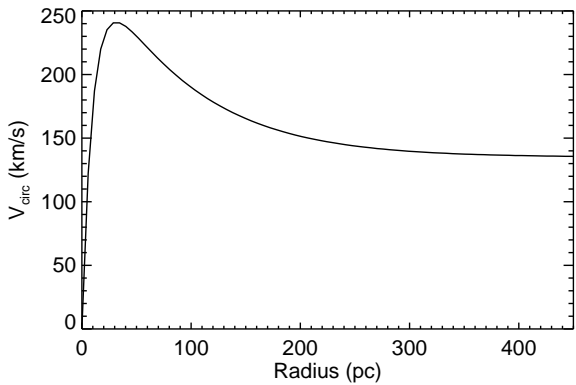}
\caption{Rotation curve adopted in our kinematic modeling of the eastern nucleus. Note: The dip interior to the peak is just there to force V = 0 at R = 0; our data do not have sensitivity to emission (or its absence) on such small spatial scales.}
\label{rotcurve}
\end{figure}

A reasonable match to the observed profiles for the eastern nucleus is
provided by a model in which the disk has center-to-edge line center
optical depth $\tau_{0,l}=4$, dust optical depth $\tau_{0,d}=1$,
velocity dispersion $\sigma = 85$ \kms, and the rotation curve is flat
with projected circular velocity $V_\mathrm{c}$ = 135 \kms. The
normalizations of the line and continuum source functions are
arbitrary; the continuum source function and the line center optical
depth have been adjusted to produce about the right levels for line
and continuum for the disk center position and to match the absorption
depth.  In order to produce the deep central absorption the line
center optical depth must be substantial ($\tau_{0,l}$ at least $\sim$
a few) and there must be a substantial gradient in the CO $J=6-5$ line
source function (i.e., the excitation temperature) as a function of
radius, decreasing outward. Thus, the deep absorption at the center is
primarily from self-absorption. However, flat rotation curve models
generically predict that the line peaks for the center position ($p=0$) lie closer to the systemic velocity than those for larger
offsets, in disagreement with the observations.  A substantial
improvement results from including a velocity perturbation (or a bump)
in the rotation curve in the inner $R\sim 100$ pc or so (compare the upper left panels of Figure \ref{model1} with the models shown in Appendix A). We have adopted the rotation curve shown in Figure
\ref{rotcurve}, in which $V_\mathrm{c}$ peaks at about 240 \kms. (The
dip interior to the peak is just there to force $V=0$ at $R=0$; we
have no sensitivity to emission (or its absence) on such small spatial
scales.) The relative positions in velocity space of the red-wing peaks are in much better agreement with the observations in the model which includes the velocity perturbation shown in the upper left panel of Figure \ref{model1} than for the flat rotation curve model shown in Appendix A (upper left panel of Figure \ref{modelprofs1}). The velocity bump
also has the effect of broadening the line profile somewhat, which
also improves the match with the data.\footnote{Although it is more
subtle, this also improves the match between the model and the data
for the $\theta_B/2$ offset positions, for the lower-amplitude peaks.}
We defer discussion of the observed rotation curve to \S 4.4.

The exact shape of the curve should not be taken too
seriously, as the needed $V(R)$ depends on the spatial distribution of
the CO line emissivity, and non-circular velocities may also play a
role. It is evident, however, that the velocity perturbation needs to
be quite significant ($\sim 100$ \kms) --- this is largely the result
of the projected velocity $V^\mathrm{proj}_\mathrm{circ} \rightarrow
0$ as the impact parameter $p\rightarrow 0$. In the context of our models, it implies that there is a significant increase in the concentration of mass interior to $R\sim 100$ pc. 

\begin{figure}[ht]
\includegraphics[scale=0.34]{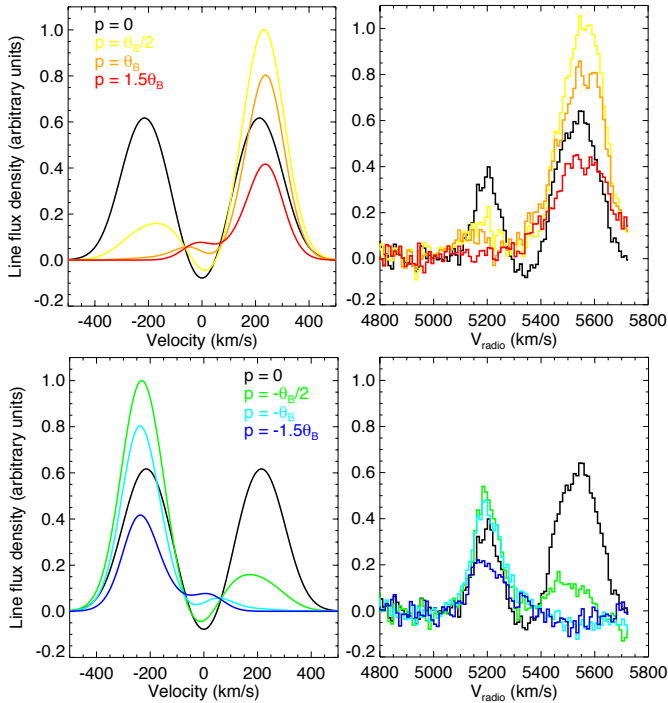}
%\vspace{+0.04in}
\caption{Comparison between the velocity profiles generated from an
  axisymmetric disk model and observations for the NE (top panels) and SW (bottom panels) sides of
  the eastern nucleus disk along the major axis of rotation. Black,
  yellow, orange and red (green, cyan, and blue) are for offsets of 0,
  $\theta_B/2$, $\theta_B$, and $1.5\theta_B$ away from the continuum
  center along the major axis approximately towards the NE (SW) side
  of the disk.}
\label{model1}
\end{figure}

The model and observed profiles are shown in Figure \ref{model1}. The
``red'' and ``blue'' sides refer to which side of the double-sided
profile dominates, so $V >0$ \kms\ (model) or $V > 5350$ \kms\
(observed) for the red side --- this is roughly the NE side of the
disk (top panel), hence, the blue side is the SW side of the disk
(bottom panel). The black line is for the central position (the
continuum peak), while the yellow, orange, and red (green, cyan, blue)
lines are for beam offsets $\theta_B/2$, $\theta_B$, and $1.5\theta_B$
along the major axis for the red (blue) sides of the disk.

The model makes a number of generic predictions that can be usefully
compared with the observed spectra along the disk major axis:

\begin{enumerate}
\item The disk center is where we see a symmetric, double-peaked line
  profile.
\item The blue/red line peak brightnesses occur away from the center
  position.
\item The blue/red line peak brightnesses and the extent of the disk
  should be symmetric about the disk center.
\item Emission in the blue wing is still present at a low level as one
  moves redward of the disk center, and vice versa.
\item Producing line profiles that are as broad and approximately
  Gaussian as observed requires a line-of-sight velocity dispersion
  that is comparable to the rotation velocity. See the appendix for a
  discussion of what constraints the line profiles place on
  $\sigma/V_\mathrm{rot}$. 

\end{enumerate}

Prediction (1) immediately implies that the disk center is at the
continuum peak, not at the peak of the line emission, i.e., even
though we find a 100 pc shift between the line peak and the continuum
peak, the center of the disk is located at the continuum peak. This
is what we would expect, as the dust emission is unaffected
by the disk kinematics. It also implies that the central velocity of
the deep absorption dip is the systemic velocity.

Prediction (2), which is the inevitable result for a rotating disk
with a source function gradient when the optical depth is not
extremely high along all lines of sight, also agrees with the
observations, but more so on the red (NE) side of the disk than the
blue (SW) side. The disagreement becomes much more pronounced with the
third prediction: the blue line peak maximum is considerably weaker
than the red peak, and the disk extent on the blue side is
noticeably less than that on the red side. Figure \ref{model1} shows that, while the ordering of peak flux density with position is basically correct, the observed line peak values are too small by about a factor of two compared to the model. In addition, the observed line widths of the blueshifted lines are
systematically smaller than those of the redshifted lines, also by
about a factor of two.

\begin{figure}[ht]
\includegraphics[scale=0.34]{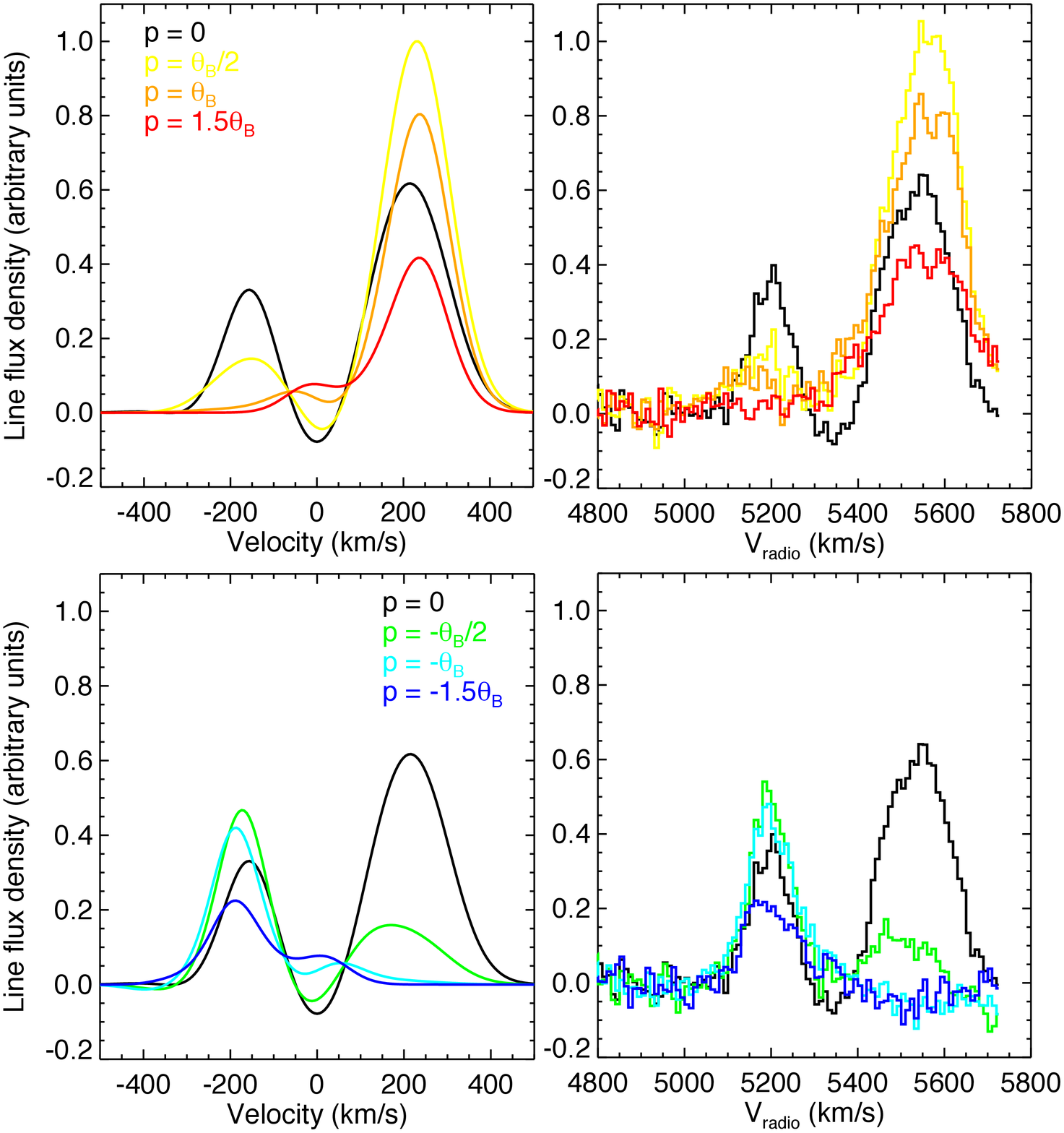}
%\vspace{-0.2in}
\caption{Comparison of the eastern nucleus kinematics between the
  axisymmetric disk plus foreground absorption model and observations
  for the SW and NE sides of the nuclear disk along the major axis of
  rotation. Color coding as in Figure \ref{model1}.}
\label{modelabs}
\end{figure}

The discrepancy with the axisymmetric model could be the result simply
of asymmetry in the disk, in excitation, optical depth (e.g., the disk
column density), or local kinematics --- the disk dynamical timescale
(radius of $100-200$ pc, rotation velocity $\sim 130$ km s$^{-1}$) is
a few million years. However, both the eastern and western nuclei show
asymmetry in the {\it same} sense: although the CO emission is more
extended than the continuum, the line emission in both nuclei is
systematically weaker on the southern side of the disks (which is the
blue side for the eastern nucleus), compared to the continuum extent
(see Figure \ref{comaps}(a,e,f)). 

There are two candidate mechanisms for producing such asymmetries in
the line emission, absorption and outflow. We consider these in turn
in the following subsections.

\subsection{Rotating disk models with foreground absorption}
The commonality of the CO line asymmetry in both nuclei suggests that something
external to the nuclei is the cause. We therefore model it as additional,
foreground absorption of the disk emission. This model assumes a Gaussian
absorption profile in velocity characterized by the line center velocity
$V_{abs}$, velocity dispersion $\sigma_{abs}$, and line center optical depth
$\tau_{abs}$, which are all allowed to vary with position. In practice,
$\sigma_{abs} = 70$ \kms\ provided a match in all cases. Because the
line/continuum level is very important in determining the optical depth, we
tweaked the model continuum levels slightly to precisely match the observations.

Consider first the blue side of the disk, where we see large
discrepancies between the original unabsorbed model and the data for
blue velocities. The absorbed model emission profiles (after continuum
subtraction) are shown in Figure \ref{modelabs}. Given the simplicity
of the model, the agreement is very good. Interestingly, there are
signs of gradients as a function of position in both the absorption
velocity and (clearly) in optical depth: the former shifts further
away from zero velocity by about 40 \kms, while the latter increases
by nearly a factor of 3 ($\tau_0 = 0.5$, $\tau_{-\theta_B/2} = 0.9$,
$\tau_{-\theta_B} = 1.3$, $\tau_{-3\theta_B/2} = 1.5$). On the red
side of the disk, only very weak absorption is allowed at
$p=\theta_B/2$, with $\tau$ no more than $\sim 0.1$, and basically
none at larger offsets. (Indeed, at $p=\theta_B$, there is clearly
emission at these velocities that is not present in the model.) Hence
in this model there is little to no absorption on the red (NE) side,
but the absorption optical depth is significant at the continuum peak
and it increases rapidly to the SW with distance from the center.

Where might this absorbing gas lie? One possibility is that both 
absorption features, i.e., the absorption that reduces the blue emission peak and the deep redshifted absorption (seen above 5505 km/s) are produced in one more-or-less contiguous feature. This would suggest that there is a very broad ($\gtrsim$650 \kms\ in velocity
space) complex of absorbing gas, the bulk of which is flowing outwards
(blueshifted with respect to the systemic velocity), assuming that
optical depth (which is probably larger for the blueshifted absorption
than for the redshifted) traces column density. The lower mass tail of
absorbing gas extending to redshifted velocities may represent
dynamical spraying of the absorbing gas over a wide range of
velocities, e.g., by tidal or ram-pressure stripping.

To explain the common CO asymmetry and redshifted absorption of the
nuclei, this gas must be at least as spatially extended as the nuclear
separation ($\sim$400 pc). Given the complicated dynamics of the
merging galaxies, the existence of such a spatial and kinematic
feature is certainly plausible. Furthermore, the spatial location of
the blue absorber would be coincident with the location of the
redshifted absorption (Figures \ref{comaps}(f)), suggesting a physical connection between the two.

Alternatively, the blue-shifted absorbing gas causing the apparent
line/continuum peak shift and the blue/red disk asymmetry could also
be explained by an outer, kiloparsec-scale molecular disk, inferred by
\citet{sakamoto09} and \citet{scoville97} from low-$J$ CO
interferometric observations. If this outer disk is suitably tilted
with respect to the two embedded nuclear disks, then the blue sides of
the nuclear disks will appear fainter than the red sides as a result
of absorption. The magnitude of the required velocity dispersion
($\sim 70$ \kms) is certainly reasonable for such a disk, especially
as it has quite likely not settled into dynamical equilibrium.

\subsection{Rotating disk models with outflow}
\citet{sakamoto09} suggested that outflows with velocity $V_{out}\sim
100$ \kms\ are present in the nuclei of Arp 220, based on their
reported detection of P Cygni profiles in the HCO$^+$~(4-3) and (3-2)
lines towards both nuclei and in the CO $J=3-2$ line for the eastern
nucleus, in SMA data with $0.3''$ resolution. Rotation is also clearly
present in their data, and their preferred model included both
rotation and outflow.

\begin{figure*}[ht]
\includegraphics[scale=0.5]{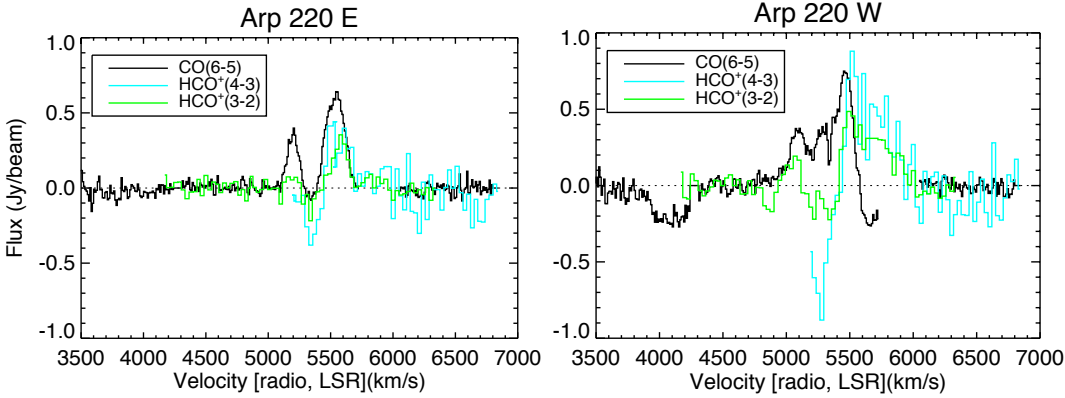}
\caption{Comparison of CO $J=6-5$ line profile (this work) with
 the HCO$^+$ (4-3)/(3-2) lines profiles from
  \citet{sakamoto09}.}
\label{saka09_comp} 
\end{figure*}

With their more limited data set, they concluded that the absorption
dip seen in the CO and HCO$^+$ lines from the eastern nucleus is
\emph{blueshifted} by $\sim 100$ \kms\ with respect to the systemic
velocity, which they estimated to be $V_{sys} = 5415 \pm 15$ \kms; a
blueshift of the absorption must be present if an outflow is
responsible for the line profile asymmetries. In Figure
\ref{saka09_comp} we overplot their line profiles on ours, for both
the eastern (left panel) and western (right panel) nucleus. For the
eastern nucleus it is clear that the velocities of the emission peaks
and of the absorption dip for all three transitions are consistent,
although the degree of asymmetry differs. We defer discussion of the
velocity shift of the absorption relative to the systemic velocity to
\S 4.4.

Since rotation is undoubtedly present, and the parameter space for
rotation plus outflow models is large, we adopted the same
axisymmetric rotating disk model as in \S 4.1. As models in which
there is a gradient in outflow velocity, rather than a constant
velocity, provided better matches, we assumed a linear increase in
outflow velocity. (Models with exponential instead of linear rises
differ insignificantly). The model outflow is symmetric about the line
of sight and has a half-opening angle $\theta_{out}$. In practice, we
found that models with $\theta_{out} \ge 30^\circ$ were
indistinguishable from models with larger half-angles, and provided
better fits than narrower outflows.

\begin{figure}[ht]
\includegraphics[scale=0.34]{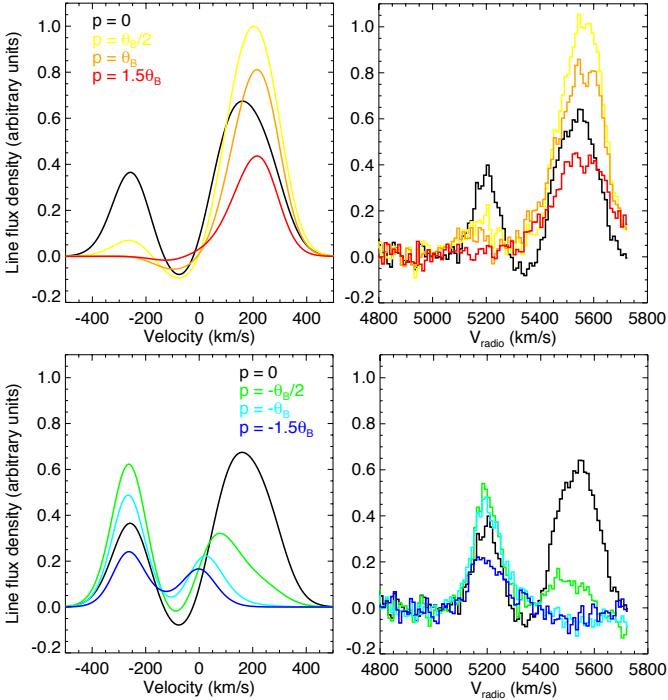}
%\vspace{-0.2in}
\caption{Comparison of the eastern nucleus kinematics between the
  axisymmetric disk plus outflow model and observations
  for the SW and NE sides of the nuclear disk along the major axis of
  rotation. Color coding as in Figure \ref{model1}.} 
\label{modelout}
\end{figure}

Reasonable approximations to the observed profiles are produced by
models with $V_{out}\simeq 100$ \kms. The resulting profiles are
compared with the observed profiles in Figure \ref{modelout}. In detail the
rotation plus outflow models do not match the data as well as the
rotation plus absorption models shown in Figure \ref{modelabs},
especially on the blue side of the disk, where they tend to produce
too much emission in the red wing of the lines. However, this
conclusion should not be considered in any way definitive, since the
amount of unexplored parameter space is very large and we
did not vary the underlying axisymmetric rotating disk model, only the
outflow parameters. Hence it is only fair to conclude that the
rotation plus outflow models for the eastern nucleus also provide an
acceptable explanation for the observed line profiles.

\subsection{The Observed and Actual Disk Velocity Curves}
\begin{figure}[ht]
\includegraphics[scale=0.37]{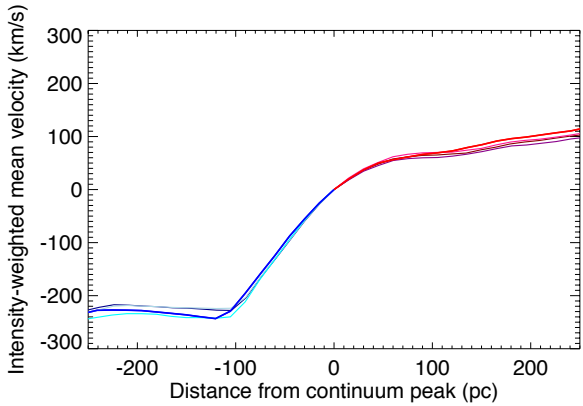}
%\vspace{-0.05in}
\caption{Observed IWMV rotation curve along the major axis of the
  eastern nucleus. The apparent systemic velocity is defined to be the IMWV at the continuum peak, 5450 \kms. We have subtracted this quantity so that the plotted curves show rotation about the apparent systemic velocity, with the approaching and receding sides of the disk color-coded appropriately. Distances are positive in the NE direction (the red side of the disk). Rotation curves for position angles
  between $28^{\circ}$ and $36^{\circ}$ are overplotted to show that the
  variations between them are extremely small.} 
\label{obsiwmv}
\end{figure}

We extract the observed rotation curve for the eastern nucleus from the moment-1
map shown in Figure \ref{comaps}(b) by taking a cut along the major axis. This
is shown in Figure \ref{obsiwmv}. The curve has been color-coded as red- and
blue-shifted with respect to the systemic velocity, which has here been defined
to be the mean velocity at the position of the continuum peak, which is 5450
\kms. To be specific, what is plotted here --- the quantity calculated by
moment-1 --- is the intensity-weighted mean velocity (IWMV) for each pixel along
the major axis. This weighting has an important influence on the relation between
the observed and actual rotation curves. Also, since the uncertainty in the mean
velocity curve is completely dominated by the uncertainty in the position angle
of the major axis, we have overplotted IWMV curves (with similar color-coding)
for position angles between $28^{\circ}$ and $36^{\circ}$, which covers the
range of uncertainty in the major axis position angle based on the continuum
emission (which is unaffected by absorption) presented in Paper I. The variation
between them is insignificant.

As mentioned in \S 4.1, the axisymmetric rotating disk model predicts that the
continuum peak is the true center of the nucleus, and the central velocity of
the self-absorption dip is the systemic velocity. This velocity is 5350
\kms\ (e.g., Figure \ref{comaps}(a)).  Hence there is a difference of
approximately 100 \kms\ from the IWMV at the continuum peak. In addition, the
observed rotation curve is noticeably asymmetric about the continuum peak
IWMV. The red side is much shallower than the blue side: at 100 pc distance from
the continuum peak, the velocity has risen by less than 100 \kms\ with respect
to systemic velocity, whereas on the blue side it is over 200 \kms.

However, the asymmetry of the observed line profiles, and in particular the
difference in profiles between the red and blue sides of the disk, produces a
systematic offset between the IWMV and the true velocity centroids of the
emission. In the presence of such line asymmetries (whatever their origin), IWMV curves can be very misleading about the underlying kinematics. As we now show, the intensity weighting of an asymmetric line profile can result in an IWMV curve which has both an offset from the true systemic velocity and an asymmetry in the derived rotation curve about the disk center, even though the underlying kinematic model is purely axisymmetric.

\begin{figure}[ht]
\includegraphics[scale=0.4]{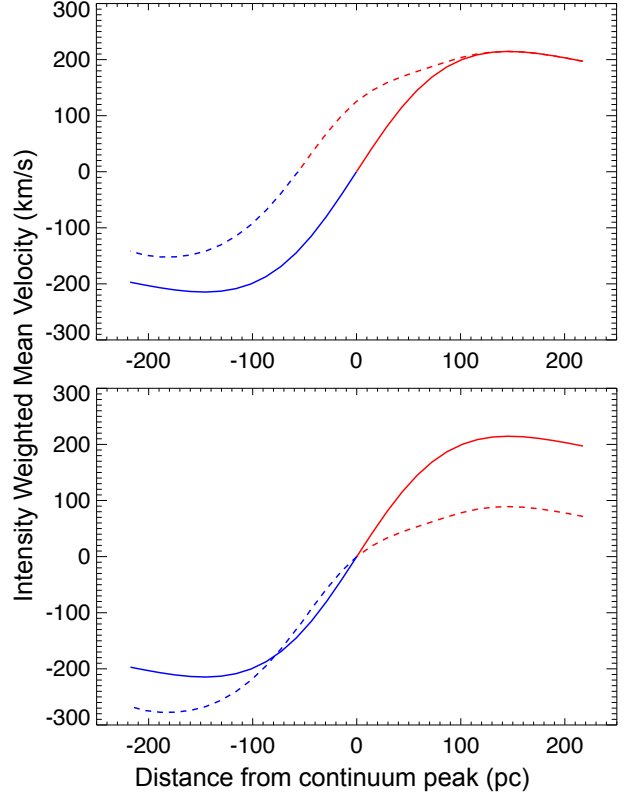}
%\vspace{-0.1in}
\caption{Rotation plus foreground absorption: Model IWMV curves, color-coded as in Figure
  \ref{obsiwmv}. \emph{Top:} Axisymmetric rotating model IWMV curve
  (solid line); with absorption (dashed line). The apparent systemic velocity is defined (as in Figure 11) as the IMWV at zero offset from the disk center. With no absorption this is at the true systemic velocity, 0 km/s. The effect of absorption is to shift this velocity to the red by $\simeq 125$ \kms. Bottom: As the top panel, except that now the apparent systemic velocity has been subtracted from the dashed curve. \emph{Bottom:} As the
  top panel, except that now the apparent systemic velocity (the IWMV at zero
  offset from the disk center) has been subtracted from the dashed curve.}
\label{modcurveabs}
\end{figure}

Figure \ref{modcurveabs} (top) shows the IWMV curve produced by the
axisymmetric model of \S 4.1, in the absence of absorption (solid
line). This is color-coded as in Figure \ref{obsiwmv}, and is of
course symmetric about the disk center, since the line profiles,
although asymmetric except for $p=0$, are mirror images of one another
on the red and blue sides of the disk. We define the apparent systemic velocity to be the IWMV at zero offset from the continuum peak (identical to the disk center), just as in Figure \ref{obsiwmv}; for the unabsorbed model this is of course equal to the true systemic velocity of 0 km/s. \footnote{It is worth noting the
weak resemblance between this ``rotation curve'' and the true disk
rotation curve shown in Figure \ref{model1}. This is a consequence of
several systematic biases, especially the $\cos\theta$ weighting of
the line-of-sight velocity component, the relatively poor spatial
resolution of the disk, the optical thickness of the lines, and the
intensity-weighting used to derive the mean velocity.} The dashed
line shows the same thing, but now for the absorbed model of \S 4.2. 
On the red side of the disk, the absorption rapidly goes to zero with
increasing distance from the disk center, so the red with-absorption
curve converges with the no-absorption curve. Since the effect of
absorption is to reduce the blue wing of emission, the absorbed IWMV
curve shifts closer to zero on the blue side, and away from zero on
the red side, with the result that the absorbed IWMV curve shifts
upward with respect to the unabsorbed IWMV curve.

However, the inferred systemic velocity is no longer zero; it has been
offset to the red, with $V_{sys} \simeq 125$ \kms. (Recall that for
the observed disk, the velocity offset between the absorption dip
central velocity, which should be the systemic velocity, and the IWMV
at the continuum peak is $\Delta V\simeq 100$ \kms\ to the red.) In the
lower panel we have subtracted off this apparent $V_{sys}$ from the
absorbed model curve. This curve now exhibits all of the features
shown by the observed IWMV curve: a steeper rise (and greater range)
on the blue side, and a shallower red IWMV curve. Hence the observed
IWMV curve is completely consistent with an underlying axisymmetric
disk model, once the effects of the line profile asymmetries are taken
into account.

\begin{figure}[ht]
\includegraphics[scale=0.4]{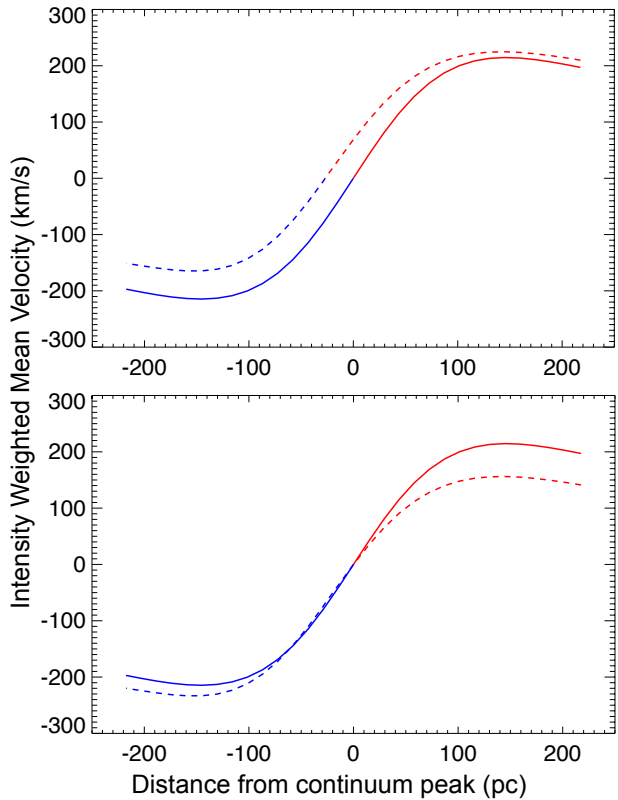}
%\vspace{-0.1in}
\caption{Rotation plus outflow: Model IWMV curves, color-coded as in Figure
  \ref{obsiwmv}. \emph{Top:} Axisymmetric rotating model IWMV curve
  (solid line); with outflow (dashed line). \emph{Bottom:} As the
  top panel, except that now the apparent systemic velocity (the IWMV at zero
  offset from the disk center) has been subtracted from the dashed curve.}
\label{modcurveout}
\end{figure}

In Figure \ref{modcurveout} we show the same pair of IWMV curves as in
Figure \ref{modcurveabs} for the outflow model of \S 4.3. Although
the overall effect of the outflow is similar to that for absorption,
the apparent red side/blue side IWMV curves are noticeably less
asymmetric than for the absorbed model, or as seen in the data. This
is because the outflow inevitably shifts line emission blueward from
where it would occur in the absence of the outflow: as seen in Figure
\ref{modelout}, the absorption dip has shifted by about 75 \kms\ to
the blue from the true systemic velocity, and so part of the red wing
of the line profile has also shifted to the blue, something which does
not happen in the absorption model. In consequence, the IWMV at the
continuum peak has shifted to the red by only about half as much as in
the absorption model (about 68 \kms\ vs 125 \kms), and the two sides of the
IWMV curves are more symmetric and less similar to the observed IMWV.

While not a definitive argument, this does suggest that outflow models
are less capable of matching the observed kinematic data than the
absorption models.

\subsection{Rotation plus foreground absorption or rotation plus outflow?}
The line profiles and kinematics of the eastern disk are fit about
equally well by a rotating, axisymmetric disk model combined with
either foreground absorption on the blue side of the disk or with an
outflow. On balance, however, we favor absorption as the explanation
for the observed asymmetries in the line profiles and the disk extent
with respect to the continuum peak:

\begin{enumerate}
\item There is unequivocally foreground absorption present at
  redshifted (with respect to systemic) velocities, over at least a
  few hundred \kms, and it occurs spatially precisely where the blue shifted absorption is expected to lie. Given the complicated dynamics and energetic
  nuclei present in Arp 220, the presence of additional absorption at
  blueshifted velocities (representing outflowing rather than
  infalling gas) would not be unexpected.
\item The line asymmetries produced by an outflow depend rather
  sensitively on the magnitude of the projected velocity compared to
  other velocities in the system, such as rotation. However, even though the two nuclear disks have very different apparent orientations, the asymmetries between the integrated line flux and the continuum emission for the two disks is very similar, as seen on the sky. This is natural in the absorption picture (the absorber only needs to be at least as large as the nuclear separation, and simply absorbs background radiation over its velocity extent) but rather contrived in the outflow model, where despite the different orientations of the nuclei, the projected rotation and outflow velocities conspire in such a way as to cause the line emission on the south sides of the disk to fade out with respect to the continuum in very similar fashion. (Recall that although we have
  not modeled the western nuclear disk, similar line profile
  asymmetries are seen there, e.g., Figure \ref{saka09_comp}.)
\item Despite the different orientations of the nuclei, the gas
  velocities on the south sides of the two nuclei (where the absorption
  occurs) are very similar (Figure \ref{comaps}(b)): from the eastern
  to the western nucleus on the blue side, all of the emission falls
  within the same range (covering $\sim 100$ \kms) so the absorbing
  gas can occupy the same range of velocities over its whole extent,
  with no need to introduce a velocity gradient across the absorber.
\item The IWMV curve produced by the rotation plus absorption model
  resembles the observed curve much more closely than the rotation
  plus outflow model, in considerable part because in the outflow
  model some of the red side emission must be shifted blueward.
\end{enumerate}

The rotation plus absorption model still
requires that there is outflowing, blueshifted gas in the Arp 220
system. However, this gas is not \emph{dynamically} responsible for
the line asymmetries. Whichever model is correct, however, the mass of
gas involved in the outflow is very substantial. In the rotation plus
outflow model, a substantial fraction of the total mass of the disk (for
which we estimate a warm gas mass $M_{\rm H_2} \simeq 1.3\times
10^8 M_\odot$, see Table 1) must be participating in the outflow in
order to produce the observed line profile asymmetries, the exact amount depending on the opening angle of the outflow. This also implies that the lifetime of the disk is only $t_{out} \sim R/V_{out} \sim
10^6$ yr.

For the rotation plus absorber model, the mass is more uncertain. We
can write the line center optical depth in a Doppler-broadened line as
\be
\tau_0 = {c^3\over 8\pi^{3/2}\nu_{i}^3} A_{ji} {g_j\over g_i} {N_i\over \Delta V}
\left(1 - e^{-h\nu_{i}/kT_{ex}}\right)
\label{tau_line}
\ee
where $\nu_i$ is the line frequency, $A_{ji}$ is the Einstein
$A$-coefficient, the $g_{i}$ and $g_{j}$ are the statistical weights of the levels,
$N_i$ is the column density in the lower state of the transition, and
$\Delta V$ is the Doppler width. For the CO $J=6-5$ line, $A_{ji} =
2.126\times 10^{-5}$ s$^{-1}$ and $\nu = 691.473$ GHz. For the
blue-side absorber, $\Delta V = \sqrt 2 \times 70$ km s$^{-1}$ and
$\tau_0 \sim 1$. The \htwo\ column density corresponding to equation
\ref{tau_line} is then
\be
N_{{\rm H_2},abs} = 2.2\times 10^{21} {Q_R\over (x_{\rm CO}/10^{-4})}
\left(1 - e^{-33.2/T_{ex}}\right)^{-1}{\rm cm}^{-2}
\ee
where $x_{\rm CO} = 10^{-4}$ is the CO abundance with respect to
H$_2$ and $Q_R = \sum g_{i} e^{-h\nu_{i}/kT_{ex}}$ is the rotational partition function for CO.

Spatially, the absorber must extend at least $\sim 400$ pc (the
separation of the nuclei) by 220 pc (the extent of the observed
absorption on the SW side of the eastern nuclear disk), giving a
minimum area $A_{abs} \approx 8.8\times 10^4$ pc$^2$. The minimum mass
of the absorber is then
\bea
M_{abs} &>& A_{abs} N_{{\rm H}_2} \mu m_{{\rm H}_2}\cr
&\approx& 4.2\times 10^6 {Q_R\over (x_{\rm CO}/10^{-4})} \left(1 -
e^{-33.2/T_{ex}}\right)^{-1}M_{\odot},
\eea
where $\mu$ is the mean molecular weight per \htwo\ molecule.

The partition function $Q_R$ is almost certainly $> 10$, and plausibly
as large as $100$, while the correction for stimulated emission is
likely $\sim 1$ to a few, giving
\be
M_{abs}\sim {\rm \ few}\times \left(10^7 - 10^8\right) M_\odot. 
\ee
This is again a very substantial amount of mass, comparable to the
total mass of warm molecular gas in Arp 220 (Note: The warm gas mass in Arp 220 is $\sim$10\% of the total gas mass). The velocity characterizing the absorber is much larger than in the outflow model, reaching blueshifted velocities of nearly 300 \kms\ with respect to systemic, as this gas must absorb the blue wings of the emission lines. 

An interesting question is whether we should expect to see the blue absorber emission away from the nuclear disks. Exploratory non-LTE modeling using RADEX \citep{tak07} of plausible absorber parameters (defined as those which produce $Q_R$ not much larger than 10, so as to minimize the mass of the absorber) suggests that the ratio of warm gas to absorber peak antenna temperatures in the $J=6-5$ line is close to a factor of 10. Furthermore the velocity extent of the absorber is almost certainly substantially smaller than that of the warm gas emission, further increasing the ratio in integrated intensity. Comparison with Figure \ref{comaps}(a) suggests that it is unlikely that the absorber emission is detectable on top of the general nuclear disk emission. There is clearly extended emission present in the channel maps between 5105 \kms\ and 5225 \kms, where we would expect the blue absorber emission to exist, but its impossible to conclude if this is from the absorbing gas rather than the general disk emission. 

ALMA observations of other CO transitions (e.g., $J=4-3, 7-6$) will be crucial in
constraining the partition function, $Q_R$, and excitation
temperature, $T_{ex}$, by providing estimates of the variation of line
optical depth with $J$, and thus reducing the uncertainty on the mass
estimate of the absorber. For example, the same absorber models described above predict that the absorber is considerably brighter in the $J=3-2$ line, both absolutely and relative to the warm gas (cf.  end of \S 3 and Figure \ref{velproflowj}).

\section{Dynamical Mass and Disk Stability}

The numbers derived from the kinematic modeling in \S 4 can be
used to estimate the enclosed mass in the eastern nucleus. Adding the asymptotic rotation
velocity (135 \kms) and velocity dispersion (85 \kms) in quadrature,
we derive a dynamical mass within $r = 220$ pc of $M_\mathrm{tot}\sim 1.6 - 2.9
\times 10^9 M_\odot$, where the larger number is obtained by
correcting the rotation velocity (but not the velocity dispersion) for
an inclination of $50^\circ$, obtained by \citet{barcosmunoz14} from a two-dimensional nonlinear least-squares fit of a
thin, tilted, exponential disk to their 33 GHz radio continuum image
with 0.08\arcsec\ x 0.06\arcsec\ resolution. Given the uncertainties,
this enclosed mass is in agreement with the total molecular gas mass
of $1.1 \times 10^{9}$ derived in Paper~I from the continuum emission
(adopting a gas-to-dust mass ratio of 100). This implies that the
dynamical mass within the radius of 220 pc is largely dominated by the
molecular gas, consistent with previous studies of Arp 220.

An interesting question is whether the nuclear disks are liable to be
gravitationally unstable to fragmentation and the formation of stars
given the large inferred mass surface densities (see Paper~I) of about
$5.4 \times 10^4 ~$\msun/$\mathrm{pc}^2$ and $1.4 \times
10^5~$\msun/$\mathrm{pc}^2$ associated with the disks of the eastern
and western nuclei, respectively. We investigate this using our model
for the eastern nucleus and the Toomre $Q$-parameter for a gaseous
disk \citep{toomre64}, in the form
\be 
Q \equiv {{v_s\kappa} \over {\pi G \Sigma}} > 1
\label{qtoomre}
\ee
for stability, where $v_s$ is the sound speed in the gas, $\kappa$
is the epicyclic frequency, $G$ is the gravitational constant, and
$\Sigma$ is the gas surface density.\footnote{Strictly speaking, this
equation applies to a thin disk, which is not the case here, since the
ratio of the velocity dispersion to the rotation velocity
$\sigma/V_c\approx 0.5-1.$ The effect of this will be to make the disk
more stable than implied by equation (\ref{qtoomre}).}

To evaluate $\Sigma$, we use our dynamical mass estimate $M_d\sim 2\times
10^9\msol$ and an effective disk area $A = \pi \theta_E^2$, where $\theta_E =
276$ pc is the deconvolved FWHM for the eastern nucleus. This gives $\Sigma
\approx 8400\msol$ pc$^{-2}$. Note that this is an average surface density (and thus lower) compared to the value reported in paper I, which was effectively the peak surface density calculated on a smaller spatial scale (radius $<$ 50~pc). We also assume a flat rotation curve with
$V_c=135$ km s$^{-1}$, so that $\kappa = \sqrt 2 \Omega = \sqrt 2 V/R$, which we
evaluate at $R=\theta_E/2$ as a typical value. Finally, we use the inferred
velocity dispersion $\sigma = 85$ km s$^{-1}$ as an effective sound speed. With
these parameters, we find $Q\sim 1.5$; this number should be regarded as
uncertain by at least a factor of two (furthermore, $\Sigma$, $V$, and $\sigma$
are all functions of radius). However, the estimated Toomre Q-parameter for the gas disk of the eastern nucleus is of order unity, suggesting that the disk is marginally stabilized against gravitational fragmentation and collapse by the large velocity dispersion. 
This large velocity dispersion requires energy input
to sustain it unless we are catching Arp 220 in a very short-lived
phase. The amount of random (non-rotational) kinetic energy in the ISM
can be written
\be
E_{kin} \sim {1\over 2} M\sigma^2 \sim 10^{56} \left({M\over 10^9
  M_\odot}\right) \sigma^2_{100}\;{\rm erg}
\ee
where $M$ is the mass of molecular gas and $\sigma_{100}$ is the
velocity dispersion in units of 100 km s$^{-1}$. Dimensionally, the
energy dissipation rate is \citep{maloney99}
\bea
\dot E_{kin}&\sim& \eta {M\over R} \sigma^3 \cr
&\sim& 6\times 10^{42} \eta \left({M\over 10^9 M_\odot}\right)
R_{100}^{-1} \sigma^3_{100}\;{\rm erg\;s^{-1}}
\label{edot}
\eea
where $R=100R_{100}$ pc is a characteristic radius and the coefficient
$\eta \lesssim 1$ \citep{maclow99}. The ratio of these two equations
gives a dissipation time
\be
t_{diss} = {E_{kin}\over \dot E_{kin}}\sim 5\times 10^5 {R_{100}\over
  \eta \sigma_{100}}\;{\rm yr.}
\ee
(A very similar estimate is obtained simply from assuming that the
areal filling factor is roughly unity and that collisions are
strongly dissipative.) For the eastern nucleus of Arp 220,
$R_{100}\sim \sigma_{100}\sim 1$, which gives the dissipation time
of only $t_{diss}\sim 10^6$ yr. This short timescale implies that there must be an energy input sustaining the large velocity dispersion. 

\section{Warm Molecular Gas Mass and  Luminosity }
We determined the mass of the warm molecular gas (listed in Table 1)
in the individual nuclei using the information from our
\textit{Herschel} SPIRE-FTS observations of Arp 220
\citep{rangwala2011}. We derived the ratio of the total warm molecular
gas mass to the total CO J = 6-5 line luminosity in Arp 220; the
former is derived from non-LTE modeling of the full CO SLED from $J =
1-0$ to $J = 13-12$ and the latter is measured directly from
observations. This ratio $M$(\htwo)/\lco(6-5) is about 21
\msun/\lsun. Using this ratio and the individual integrated CO $J=6-5$
ALMA fluxes for the two nuclei we estimate the warm molecular gas mass
to be $1.4 \times 10^{8}$ \msun\ and $1.6 \times 10^{8}$ \msun, in the eastern and
western nucleus, respectively.\footnote{We note that the beam size of
\textit{Herschel} is large compared to the angular extent of Arp~220 and therefore this ratio is an average
quantity over the whole system and is likely to vary spatially. Hence,
our mass estimate for the warm molecular gas for the individual nuclei
should be used with caution.} These are roughly 1/10th of the total
molecular gas masses measured from dust continuum emission in Paper I
and are consistent with the warm molecular gas mass fraction found in
nearby galaxies \citep{kamenetzky14}.

We list the observed $L_{\mathrm{CO J=6-5}}$/\lfir\ ratio in the two
nuclei in Table 1 using the \lfir\ value calculated in Paper I. This
ratio differs by almost an order of magnitude between the two nuclei
-- a direct result of their \lfir\ values differing by an order of
magnitude. The ratio in the eastern nucleus of $\sim6.7 \times 10^{-5}$
is much higher than the western nucleus; the value for the latter is
within a factor of two of the average value of $\sim 10^{-5}$ found in
nearby galaxies \citep{kamenetzky14}. Despite the disparity in the
ratio of CO $J=6-5$ luminosity to bolometric luminosity, the
line-to-continuum ratios at 691 GHz (listed in Table 1) are the same
for the two nuclei, implying that the eastern nucleus contains an
additional power source which is depositing energy into the gas
without affecting the dust continuum emission. Such an energy source
is consistent with mechanical energy from shocks. A similar scenario
is inferred in another nearby IR-luminous galaxy, NGC 6240, in which
the CO is also unusually luminous relative to the bolometric
luminosity, with $L_{\mathrm{CO J=6-5}}$/\lfir\ $\sim7.6 \times
10^{-5}$ \citep{meijerink13}. It is intriguing that even though there
is evidence of large-scale gas dynamics (both outflow and inflow) towards both nuclei, only the eastern nucleus -- which is the less luminous of the two, and which has not been suggested to host an AGN -- exhibits enhanced CO luminosity relative to its FIR luminosity.

The large velocity dispersion found for the eastern nucleus is one
possible source of energy for the warm CO emission. From
\citet{rangwala2011} \citep[also,] []{kamenetzky14}, the total CO
luminosity of Arp 220 is $L_{\rm CO}\approx 2\times 10^8
L_{\odot}$. We assume that half of this luminosity comes from the
eastern nucleus. However, at the warm temperature inferred for this
component, most of the gas cooling will be from molecular hydrogen,
not CO \citep{neufeld95}. From \citet{bourlot99}, the H$_2$ cooling
function is $\Lambda_{\rm H_2} \approx 10^{-22}$ erg s$^{-1}$ (= 11 \lsun/\msun) with a
variation of about a factor of three over the range of temperatures
and densities that characterize the warm molecular gas in Arp
220. With a warm gas mass $M_{\rm H_2} \simeq 1.3\times 10^8 M_\odot$
for the eastern nucleus (Table 1), the total warm gas cooling $L_{\rm
warm}\sim 6.5\times 10^{42}$ erg s$^{-1}$ (with a range of a factor of
3). Comparison with equation \ref{edot} shows the energy dissipation
rate is of the right order of magnitude to excite the warm molecular
emission. This new result disagrees with the conclusions of Rangwala et al (2011). Their much smaller estimate of the energy available from turbulent heating was based on a gas velocity gradient derived from modeling of the global CO emission, while our new value uses the directly observed velocity dispersion provided by our ALMA observations. 

We also consider the energy available from supernovae and stellar winds. Using the observed supernovae rate of $\nu_{\rm SN} = 4 \pm 2$ per year \citep{lonsdale06} and ($E_{\rm SN} = 10^{51}$ erg)  in the following expression from \citet{maloney99}:
\be
L_\mathrm{SN} \sim 3 \times 10^{43} (\frac{\nu_\mathrm{SN}}{1 \mathrm{yr}^{-1}})  (\frac{E_\mathrm{SN}}{10^{51}\,\mathrm{ergs}})\, \mathrm {erg}\,\mathrm{s}^{-1}, 
\ee
we obtain a total mechanical energy of $\sim 10^{44}$ erg s$^{-1}$. The energy output from stellar winds is similar to supernovae \citep{mccray87}, giving a total of $\sim 2 \times 10^{44}$ erg s$^{-1}$, also large enough to power the warm gas emission and maintain the observed velocity dispersion; this is consistent with the conclusions of \citet{rangwala2011}.

\section{Highly Excited SiO Absorption Towards the Western Nucleus}

Absorption from the SiO $J =16-15$ line at 694.293 GHz is seen towards
the location of the continuum emission from the western nucleus
(Figure~\ref{comaps}(e) and Figure \ref{velprofcen}).
The absorption is point-like with a full velocity width of at
least 400 km s$^{-1}$; the line width could be as large as 500 km
s$^{-1}$ if the SiO absorption continues through the small gap in our
spectral coverage. The mean depth of the absorption
in the central 200 km s$^{-1}$ of the line is -0.23 Jy beam$^{-1}$ or -6.9 K.
No significant SiO absorption is seen towards the eastern nucleus;
any absorption in this nucleus must be at least 12 times fainter over
the same relative velocity range.

The SiO $J = 5-4$ line at 217 GHz has been previously seen in Arp~220 in emission
\citep{martin11}, although the angular resolution of the data was
insufficient to distinguish between the two nuclei. \citet{martin11}
note that the SiO emission is not particularly enhanced in Arp 220
compared to other species. In contrast, \citet{garciaburillo10} find
that SiO emission near the AGN in NGC 1068 has an enhanced abundance
by a factor of 10-100 compared to locations in the starburst ring. The
fact that \citet{martin11} do not see an enhanced SiO abundance may be
related to the inability of their data to isolate the region nearest
to the AGN from surrounding starburst regions.

For Galactic star forming regions, \citet{schilke97} showed that SiO
is released via neutral particles colliding with charged grains. Shock
speeds of 25 \kms\ and densities of $1\times 10^5$ \cmthree\ are
sufficient to reproduce the observed SiO column densities. The mean
density of the dusty disk in the western nucleus of Arp 220 is
estimated to be $\gtrsim 10^5$ \cmthree\ (Paper~I) and high turbulent
velocities would be consistent with the large CO line-widths seen
towards this nucleus. Thus, we might expect significant SiO to be
present throughout this dense dusty disk.  However, the SiO J=16-15
absorption is clearly more compact than the dusty disk; the strength
of the SiO absorption does not change as the beam is increased to a
FWHM of 0.5$^{\prime\prime}$, while the peak continuum flux increases
from 1.15 to 1.69 Jy beam$^{-1}$. Thus, high excitation SiO molecules
appear to be concentrated towards the center of the western
nucleus. However, because the dust emission is quite optically thick
($\tau \sim 5$, Wilson et al. 2014), the radius of the region
producing the SiO absorption has to be comparable to the scale height
of the disk.

The question arises as to how SiO is excited to such a high energy
level in Arp 220.  The upper state of the SiO $J = 16-15$ line is 283
K above ground and the transition has an Einstein coefficient $\log
A_{ij} = -1.74$ (taken from the Splatalogue molecular database). Thus, the
critical density of this transition is $\sim 2 \times 10^8$ cm$^{-3}$,
1000 times higher than that of the CO $J = 6-5$ line. Because the SiO
$J = 16-15$ line is seen in absorption, the excitation temperature of
this transition must be $< 100 - 200$~K, the estimated brightness
temperature of the continuum source at this frequency.  One possible
mechanism to excite SiO is via radiative pumping. Radiative pumping
has been invoked to explain the high-J absorption lines of HCN seen in
Arp 220 \citep{rangwala2011}.  The first excited vibrational state of
SiO is at $\sim8$ $\mu$m, near the peak of a 360 K blackbody
spectrum. Given the high mean temperature of the western nucleus,
radiative pumping by the intense radiation field of an AGN or an
extreme nuclear starburst appears to be a viable mechanism.

The alternative would be collisional excitation in the inner,
presumably higher density regions, of the nuclear disk. However, a 50
pc radius disk with a mean density of $4 \times 10^5$ cm$^{-3}$
requires a steep density profile ($\sim r^{-1})$ to achieve densities
comparable to the critical density before the dust reaches its
sublimation temperature. Therefore, it is difficult to see how a
sufficiently extended region of excited SiO could be achieved under
these conditions.

\section{Conclusions}
Our Cycle-0 ALMA observations of the CO $J=6-5$ line show complex features in
the morphology and kinematics of the warm molecular gas not seen before, neither
in low-$J$ lines tracing cold gas nor in lower-resolution CO $J=6-5$
observations from the SMA. The overall morphology of the warm emission is
similar to that traced by the low-$J$ lines. However, we detect an apparent
offset of $\sim$100 pc between the peak of the continuum emission and the peak
of the line emission. This offset is present in both nuclei and is in the same
direction. Comparison with kinematic models suggests that the offset
arises due to the absorption of the entire South (blue) side of the central region
containing the two nuclei by gas lying in front of the nuclei. There is also
clear evidence in our data for redshifted absorption, which could be interpreted as an inflowing molecular filament. If a single coherent feature is responsible for all
of the absorption features - both the redshifted absorption and the blueshifted
absorption that produces the apparent offset and the line profile asymmetries -
it has a total velocity extent of $\sim$700 \kms\ and a spatial extent of
$\sim$400 pc. Alternatively, the blueshifted absorption may arise in a larger
(kpc)-scale molecular disk, as suggested by earlier interferometric
observations, and have no dynamical connection with the infalling gas. 
Although similar asymmetries in the line profiles can be produced by a
combination of rotation plus outflow rather than rotation plus absorption, on
balance we believe that the data favor absorption rather than outflow as the
source of the asymmetries. 

From our modeling, we conclude that the dynamical center of the eastern nuclear
disk is at the location of the continuum peak, and that a large gradient in the
excitation is needed to explain the deep absorption detected at the center of
the eastern nucleus. In addition, the velocity dispersion in the disk must be a
significant fraction of the rotation velocity to be consistent with the large
line widths observed in the two nuclei. The dynamical mass estimated from our
kinematic model implies a very large mass surface density for the eastern
nucleus, suggesting the disk could be gravitationally unstable and prone to
disruption. However, the estimated Toomre $Q$-parameter for the gas disk of the
eastern nucleus is (marginally) greater than unity, suggesting that the disk is
plausibly stabilized against gravitational fragmentation and collapse by the
large velocity dispersion.

We measure an unusually high ratio of CO $J=6-5$ luminosity to the total FIR
luminosity in the eastern nucleus compared to the western nucleus; for the
latter the value is consistent with the average found in nearby galaxies. This
suggests that there is an additional energy source, such as mechanical energy
from shocks, present in the eastern nucleus that is generating more CO
luminosity without affecting the dust continuum emission and hence the total FIR
luminosity. We find that the total warm gas cooling from \htwo\ approximately matches the energy dissipation rate (from velocity dispersion), suggesting that this energy dissipation could also potentially power the warm molecular emission, in addition to the energy from supernovae and stellar winds.

Without the exceptional sensitivity and spatial/spectral resolution from ALMA it
would have not been possible to detect the complexity in the morphology and
kinematics of the warm molecular gas in Arp 220, and to measure the various
quantities associated with them with high signal-to-noise ratios. This is especially
important for detecting the signature of inflowing molecular gas, which is much
harder to detect. The high S/N of finely sampled velocity profiles allowed us
to distinguish between various kinematic models. It will be highly beneficial to
have the same quality of observations for the multiple CO lines that are accessible
with ALMA, to look for spatial variations in the gas excitation and better characterize the mass and excitation of the large-scale gas dynamics (both outflow and inflow) towards the two nuclei. This kind of spatial/kinematic analysis is highly complementary to the analysis of Herschel observations described in \S 1. Our accepted Cycle-2 observations of Arp 220 (and NGC 6240) will provide a step
forward in this direction.

\acknowledgments 
The National Radio Astronomy Observatory is a facility of the National
Science Foundation operated under cooperative agreement by Associated
Universities, Inc. This paper makes use of the following ALMA data:
ADS/JAO.ALMA\*2011.0.00403.S. ALMA is a partnership of ESO
(representing its member states), NSF (USA) and NINS (Japan), together
with NRC (Canada) and NSC and ASIAA (Taiwan), in cooperation with the
Republic of Chile. The Joint ALMA Observatory is operated by ESO,
AUI/NRAO and NAOJ.
We are very grateful to Adam Leroy for doing a custom reduction for our observations and providing very useful recommendations. 
We thank Kazushi Sakamoto for sharing his CO $J=3-2$ data cube observed by the Submillimeter Array. 
The research of Naseem Rangwala is supported by NASA ROSES grant NNX13AL16G.
The research of Philip R. Maloney is supported by NASA grant 1487846.
The research of Christine D. Wilson (C.D.W.) is supported by grants from the Natural
Sciences and Engineering Research Council of Canada. C.D.W. also
thanks the European Southern Observatory and the National Radio
Astronomy Observatory for visitor support. 
The research of Jason Glenn is supported by NASA grant 1472566.

\newpage
\appendix
\section{Appendix A: Additional Modeling Results}

\begin{figure}[ht]
\includegraphics[scale=0.35]{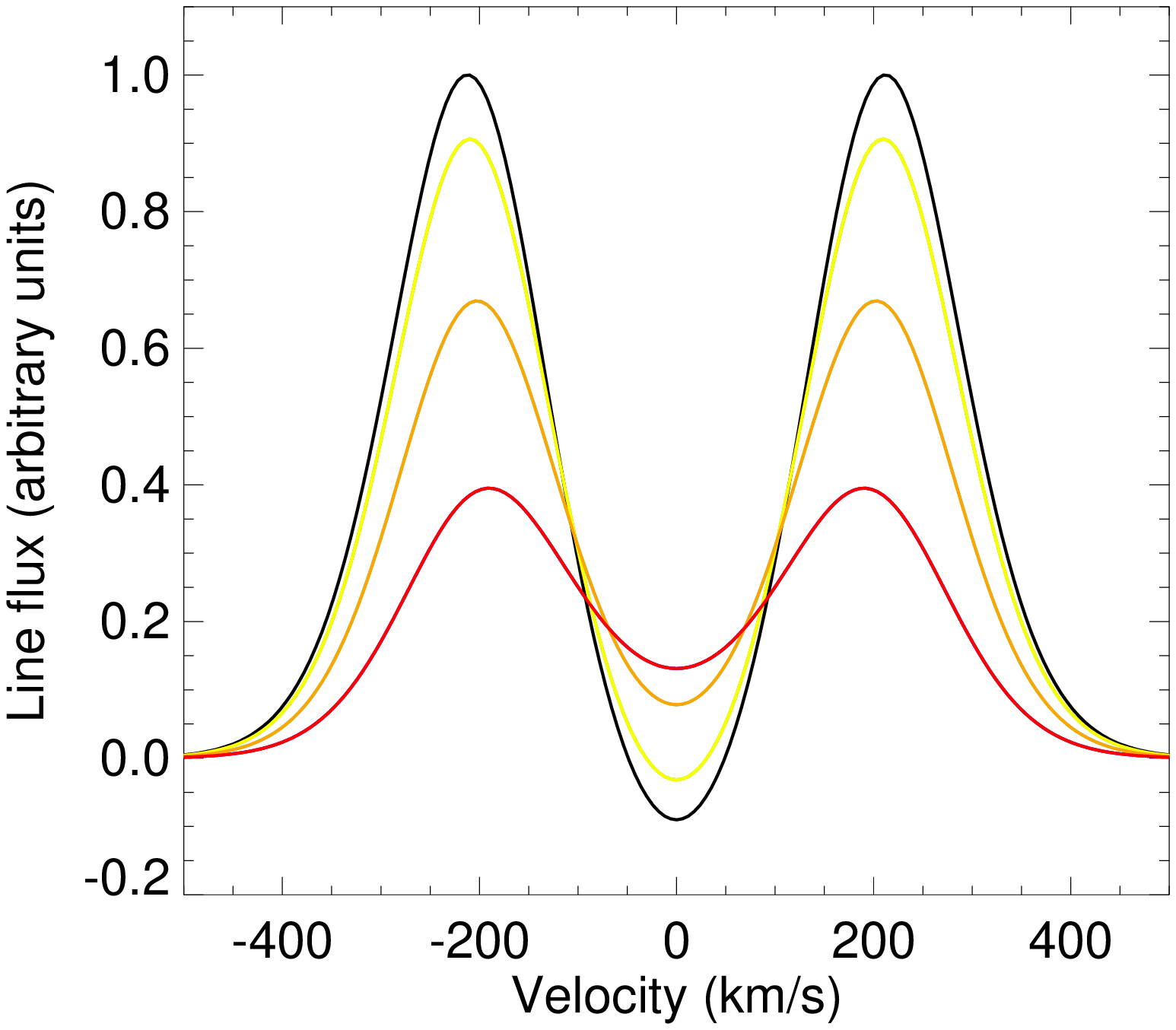}
\includegraphics[scale=0.35]{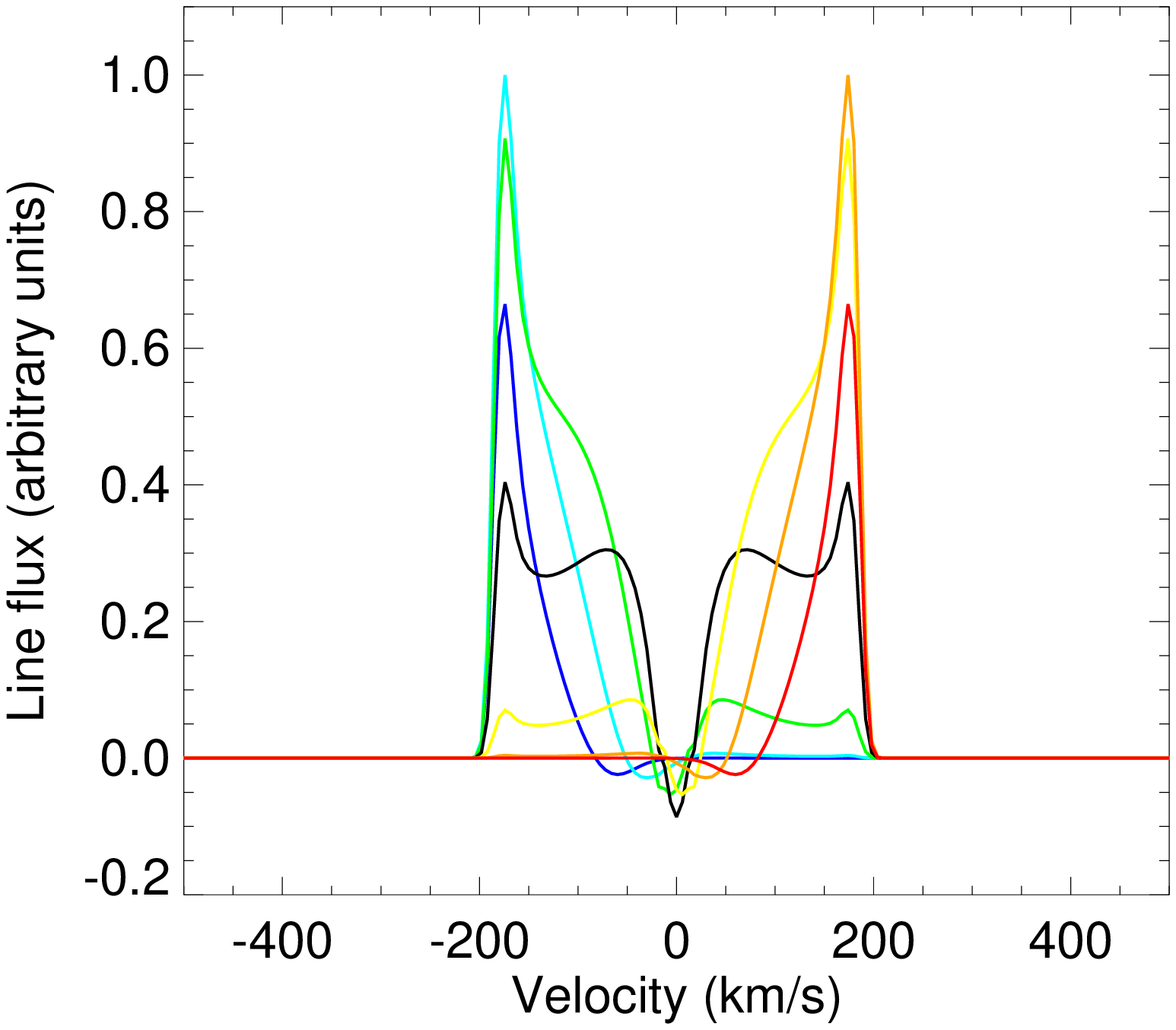}
%\vspace{-0.2in}
\caption{Velocity profiles generated from an axisymmetric disk model
  for two cases: (Top) Doppler parameter ($\sqrt{2}$ $\times$ dispersion) = 180 \kms\ and Vmax = 0
  \kms\ and (Bottom) Doppler Parameter = 10 \kms\ and Vmax = 180 \kms. Color coding as in Figure \ref{model1}.} 
\label{modelprofs2}
\end{figure}

\begin{figure}[ht]
\includegraphics[scale=0.35]{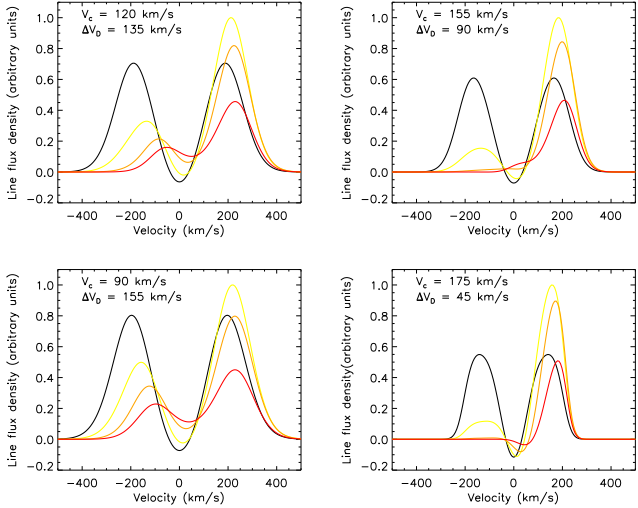}
%\vspace{-0.2in}
\caption{Model profiles for different values of Doppler parameter
  and maximum rotation velocity with their  quadrature sum fixed. Color coding as in Figure \ref{model1}.}     
\label{modelprofs1}
\end{figure}

The strength of the emission on the opposite-velocity side of the
absorption dip as one moves away from the center along the major axis
towards either the red or blue sides of the disk, the relative peak
heights along the major axis, and the shapes and velocity extents of
the red and blue line profiles, can be used as diagnostics of the
relative magnitudes of the velocity dispersion and rotation
velocity. Since there are clearly systematic issues with the
blue side of the disk, discussed in Section 4, we mostly consider the
constraints provided by the red side of the emission. In the limit of
zero rotation velocity (assuming the same optical depth and source
function parameters as in Figure \ref{model1}), the line profile is
symmetric about the absorption dip, and the line peak flux densities
are greatest at the disk center, while if the velocity dispersion is
much smaller than the rotation velocity, the line profiles are very
non-Gaussian, and peak sharply at the maximum rotation velocity at all
positions along the major axis (see Figure \ref{modelprofs2}). To illustrate the approximate allowed
ranges of the rotation velocity and velocity dispersion, in Figure
\ref{modelprofs1} we plot the resulting line profiles (at
half-beamwidth intervals and Gaussian-weighted, as in Figures
\ref{model1}) for four different models, each labeled by $V_c$ and
$\Delta V_D$. In all cases $V_c$ and $\Delta V_D$ have been varied so
that their quadrature sum is nearly fixed, which ensures that the
peak-to-peak separation of the red and blue peaks is similar to the
observed ($\sim 350$ \kms) in all models. Note that this
separation is generally significantly larger than twice the rotation
velocity, as is the extent of the emission in velocity space: this is
a consequence of both the substantial velocity dispersion and of
optical depth effects, as the models are not optically thin. For
clarity, we only plot the models for the center and red-side
positions; the blue-side emission lines are the mirror-image of the
red-side.

There is little change in the peak flux densities as $\Delta V_D$ is
increased to 135 \kms; the main change is a modest increase in the
line widths. As is evident in Figure \ref{model1}, the model line profiles are a
little narrower than the observed, and do not exhibit the trend of
increasing linewidth with distance along the major axis seen at the
$\theta_B$ and $1.5\theta_B$ offset points, indicating deviations from
the model assumptions (constant $V_c$ and $\Delta V_D$). It does
predict a slightly stronger red peak at $\Delta\theta = -\theta_B/2$
(the blue side of the major axis) than the $\Delta V_D = 120$ \kms\
model, which is already slightly too strong compared to the
observations.

Raising the rotation velocity to 155 \kms\ makes the lines narrower,
slightly reduces the difference in peak flux density between the
center and $\Delta\theta = 3\theta_B/2$ positions, and substantially
reduces the intensity of the blue-shifted emission. Reducing $V_c$ to
90 \kms\ and raising $\Delta V_D$ correspondingly (i.e., swapping the
values from the previous model) makes the center and $\Delta\theta =
\theta_B$ peak flux densities nearly equal, in substantial
disagreement with the observations, and makes the blue-shifted
emission substantially brighter. Finally, raising $V_c$ as high as 175
\kms\ produces nearly equal peak flux densities in the center and
$\Delta\theta = 3\theta_B/2$ positions, and more nearly equal peak
flux densities in the $\Delta\theta = \theta_B/2$ and $\Delta\theta =
\theta_B/$ positions, again in disagreement with the observations. 

In summary, these model variations suggest that the disk rotation
velocity is unlikely to be substantially below 120 \kms, or above 155 \kms, with the velocity dispersion scaled accordingly. 

\bibliographystyle{apj}

%\bibliography{nsf}
\end{document}